\documentclass[]{article}
\usepackage{amssymb,amsfonts,amsthm,bm,graphicx,amsmath,float,color,longtable}
\restylefloat{figure}
\usepackage[ansinew]{inputenc}
\usepackage{authblk}

\newcommand\blfootnote[1]{%
  \begingroup
  \renewcommand\thefootnote{}\footnote{#1}%
  \addtocounter{footnote}{-1}%
  \endgroup
}

\usepackage{crop}%
\usepackage{amsmath,amssymb,amsfonts,upref,endnotes}
\usepackage{amsthm}
\usepackage{marginnote}
\usepackage{soul}
\RequirePackage{graphicx}
\usepackage[figuresright]{rotating}
\usepackage{color}

\usepackage{fancyhdr}
\usepackage{lastpage}
\usepackage[scaled]{helvet}
\usepackage[T1]{fontenc}

\definecolor{jobcolor}{cmyk}{1,.81,.09,.01}
\usepackage{xcolor}
\usepackage{sectsty}
\sectionfont{\color{jobcolor}}  

\usepackage{helvet,times}
\usepackage{bm,textcomp}
\usepackage{amsmath,amsthm,amssymb,amsbsy,epsfig,fancyhdr,calc,ifthen,float,psfrag,mathrsfs,graphics,color,fullpage}
\restylefloat{figure}
\usepackage{subcaption} 
\usepackage[ansinew]{inputenc}

\usepackage{empheq}

\newcommand*\widefbox[1]{\fbox{\hspace{2em}#1\hspace{2em}}}

\begin{document}

\title{Mechanics of torque generation in the bacterial flagellar motor}

\author[1]{Kranthi K. Mandadapu$^{*}$}
\author[2]{Jasmine A. Nirody$^{*}$}
\author[3]{Richard M. Berry}
\author[4]{George Oster$^{\dagger}$}
\affil[1]{Department of Chemistry, University of California, Berkeley}
\affil[2]{Biophysics Graduate Group, University of California, Berkeley}
\affil[3]{Department of Physics, University of Oxford, Oxford, UK}
\affil[4]{Department of Molecular and Cellular Biology, University of California, Berkeley}
\date{}

\maketitle

\begin{abstract}
{The bacterial flagellar motor (BFM) is responsible for driving bacterial locomotion and chemotaxis, fundamental processes in pathogenesis and biofilm formation. In the BFM, torque is generated at the interface between transmembrane proteins (stators) and a rotor. It is well-established that the passage of ions down a transmembrane gradient through the stator complex provides the energy needed for torque generation. However, the physics involved in this energy conversion remain poorly understood. Here we propose a mechanically specific model for torque generation in the BFM. In particular, we identify two fundamental forces involved in torque generation: electrostatic and steric. We propose that electrostatic forces serve to position the stator, while steric forces comprise the actual `power stroke'. Specifically, we predict that ion-induced conformational changes about a proline `hinge' residue in a $\alpha$-helix of the stator are directly responsible for generating the power stroke. Our model predictions fit well with recent experiments on a single-stator motor. Furthermore, we propose several experiments to elucidate the torque-speed relationship in motors where the number of stators may not be constant. The proposed model provides a mechanical explanation for several fundamental features of the flagellar motor, including: torque-speed and speed-ion motive force relationships, backstepping, variation in step sizes, and the puzzle of swarming experiments.}
\end{abstract}
\blfootnote{$^{*}$ These authors contributed equally.}
\blfootnote{$^{\dagger}$ Corresponding author: goster@berkeley.edu}
The bacterial flagellar motor (BFM) is one of only two known protein motors that utilizes the potential energy stored in the transmembrane ion gradient (the \emph{ion motive force}, or IMF) instead of ATP, the near-universal cellular energy currency. The other such motor is the one responsible for the synthesis of this energy source---the F$_\text{O}$ motor of ATP synthase. Understanding how these ion-driven machines generate useful mechanical work is a fundamental issue in cellular biology. Accordingly, much work has been devoted to understanding the torque-generation mechanisms of these molecular machines.

One of the principle diagnostics for a rotary protein motor is the relationship between motor torque and rotational speed. Theoretical models attempt to reproduce these empirically measured relationships. The torque-speed curve of the BFM was famously shown to display two distinct regimes: a constant-torque plateau at low speeds that sharply transitions into a near-linear decrease in torque at high speeds \cite{Berg2003}. However, recent experiments from the Berg laboratory show that the number of torque-generating units (or \emph{stators}) is likely not constant across this curve \cite{Lele2013}. That is, stators were shown to come on and offline as the applied load was varied. This is akin to a car in which the number of active cylinders changes as the car goes up and downhill. This finding effectively invalidates all current theoretical models of torque-generation in the BFM, as they have been based on this experimentally measured curve and assumed that the number of working stators is constant \cite{yuan2010asymmetry, xing2006torque, Meacci2009, Meacci2011}.

Recent torque-speed measurements have been performed on single-stator motors \cite{lo2013mechanism}, providing insight into the intrinsic torque-generation mechanism of the BFM and making this theoretical reexamination especially timely. Furthermore, current models describe torque generation phenomenologically as a free energy surface without committing to a specific physical mechanism. In the following we combine the currently available structural information with biophysical and biochemical studies on the dynamical behavior of the motor. We use this compiled information to propose a \emph{mechanically specific} and \emph{experimentally testable} model of torque generation in the BFM.
\begin{figure}
\begin{center}
\centerline{\includegraphics[width=8.7cm]{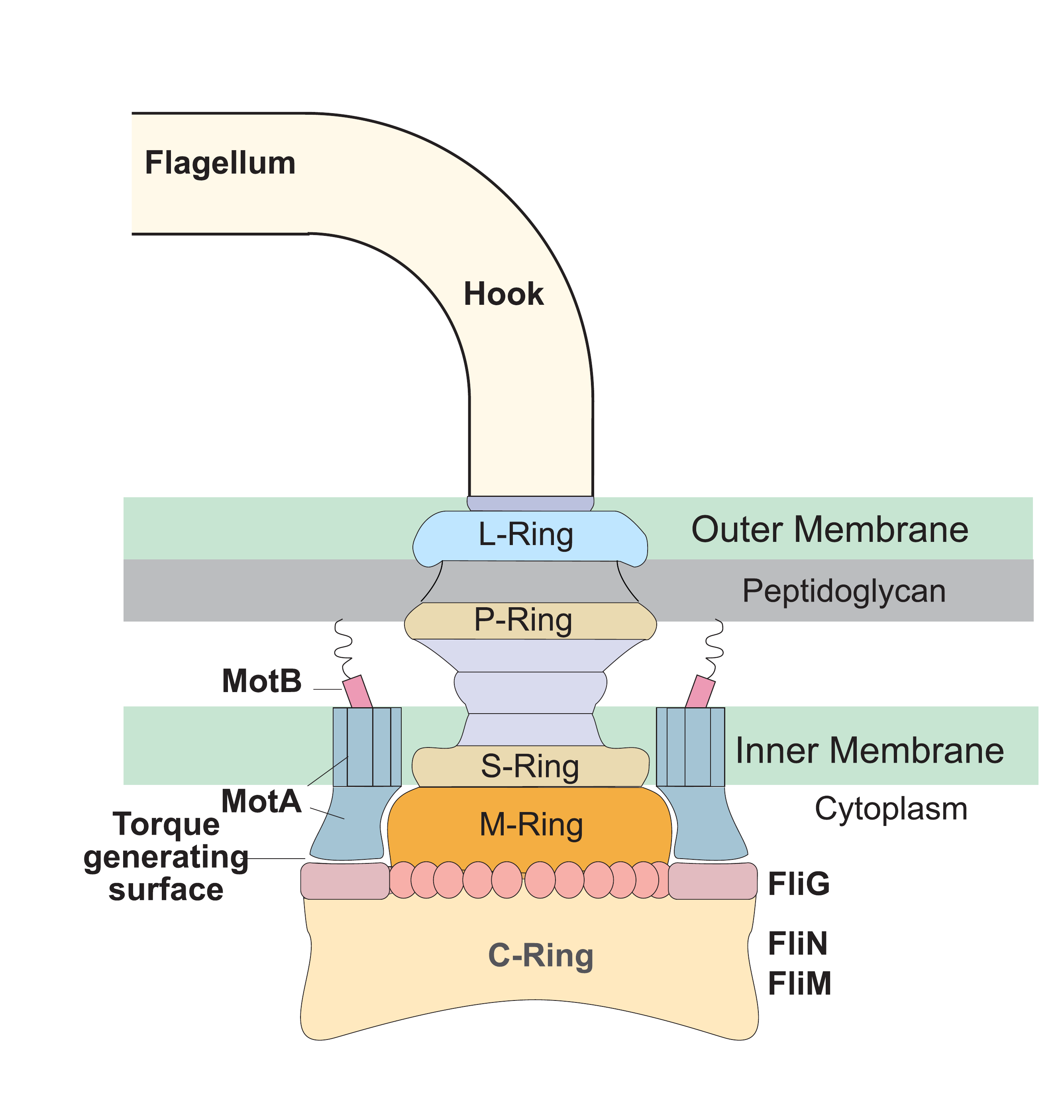}}
\label{fig:deepthi}
\caption{Schematic showing the basic parts of the BFM. A bacterium has on average four flagellae, each attached to the basal body of a motor via a flexible hook. The M-, S-, and C-rings of the basal body are together called the `rotor'. FliG proteins (26 copies of which are assumed here) are placed around the periphery of the C-ring. These interact with the MotA loops of the `stator' to generate torque and rotate the flagella. Stators are composed of MotA and MotB subunits, the latter of which attaches the stators to the peptidoglycan layer, allowing for torque generation via the MotA-FliG interaction. A motor can have between 1 and 11 engaged stators, depending on the load \cite{Lele2013,tipping2013load}.}
\end{center}
\end{figure}
The BFM consists of a series of concentric rings embedded in the cell envelope connected to an extracellular helical propeller by a flexible hook (Figure~1). The cytoplasmic \textit{C-ring} acts as the rotor and the membrane-embedded \textit{Mot} complex(es) act as the stator(s). A motor can have between 1 and 11 such stator units. Each stator unit is composed of 4 MotA and 2 MotB helix bundles \cite{braun2004arrangement, Kim2008}. A MotA bundle consists of four membrane-embedded $\alpha$-helices linked by two large cytoplasmic loops. The interaction between these cytoplasmic loops with FliG proteins located on the periphery of the C-ring is implicated in torque-generation. We note that while there is some controversy in the exact number of FliGs, this detail does not affect the main points of our model. For ease of exposition, in the following we assume that there are 26 FliG `spokes' on the rotor.

The feat of coupling of an ion gradient to the generation of mechanical work is attributed to the peptidoglycan-bound MotB complexes. These complexes each contain an ion-conducting channel with a negatively-charged aspartate residue (Asp32), which functions in cation binding. This residue is one of the only strongly conserved residues across bacterial species. We propose that the interaction between the Asp32 and a cation passing across the inner bacterial membrane (between the periplasm and the cytoplasm) induces conformational changes in the stator complex, resulting in the torque-generating `power stroke' \cite{Kim2008}. While a crystal structure of the stator complex is needed for a complete understanding of the power stroke, the available structural knowledge combined with information about the motor's dynamical performance is sufficient to propose a plausible and experimentally testable model. Recent structural information suggests that a proline residue in MotA is essential for torque generation \cite{Braun1999,Kim2008,Lee2010}. Using this information, we present a mechanical model for torque generation involving proline-induced conformational changes in MotA cytoplasmic loops \cite{Cordes2002, Kim2008}.  

To the best of our knowledge, our model is the first to incorporate known structural information about the BFM stator and rotor complexes into a quantitative physical mechanism for the generation of the power stroke. As part of this study, we aim to address the following fundamental questions: 
\begin{itemize}
\item What are the kinematics and mechanics of the BFM power stroke?
\item What role do charged residues on the stator and rotor play in torque generation, and how does this role explain mutational experiments which show only a partial reduction in motor efficiency? 
\item What is the shape of the torque-speed curve for motors with a single stator?  
\item Why does the motor exhibit backsteps even in the absence of an external `reversal' signal (usually the small protein CheY-P)? 
\end{itemize}

Here we primarily discuss the proton (H$^+$) powered motor of \textit{Escherichia coli}. However, our model is sufficiently general so as to apply to the sodium (Na$^+$) powered motors found in alkalophiles and marine \textit{Vibrio} species. 

\section*{The Mechanochemical Model}
\subsection*{An electro-steric power stroke} 

Due to the small magnitude of the forces involved relative to thermal fluctuations, it has long been assumed that nearly any form of interaction would be sufficient to explain the rotation of the BFM \cite{Berg2003}. For this reason, previous models have avoided characterizing the exact nature of these forces, instead treating the interaction between the stators and the rotor phenomenologically as a free energy surface and the stator as an \textit{ad hoc} stochastic stepper \cite{xing2006torque, meacci2009dynamics, meacci2011dynamics}. 

However, knowledge gained from recent structural \cite{brown2002crystal, braun2004arrangement,lowder2005flig} and biophysical \cite{lo2013mechanism} studies have led us to conclude that the power stroke of the BFM is \textit{electrosteric}---that is, it is driven by both electrostatic and steric forces. In the following, we propose a mechanochemical model consisting of two phases. (i) During `electrostatic steering', electrostatic forces position the stator. (ii) Once positioned, the stator delivers a steric push (i.e., a contact force) to a FliG protein located along the periphery of the rotor. A more detailed description of the nature of contact forces is found in the supplementary text and in reference \cite{happel1983low}. 

In the following we lay out the assumptions involved in the construction of our model, followed by a detailed description of the mechanism. Details of the mathematical formulation are provided in the Materials and Methods.

\subsection*{Electrostatic forces steer the stator into place}
\begin{figure}[t]
\begin{center}
\centerline{\includegraphics[width=12.5cm]{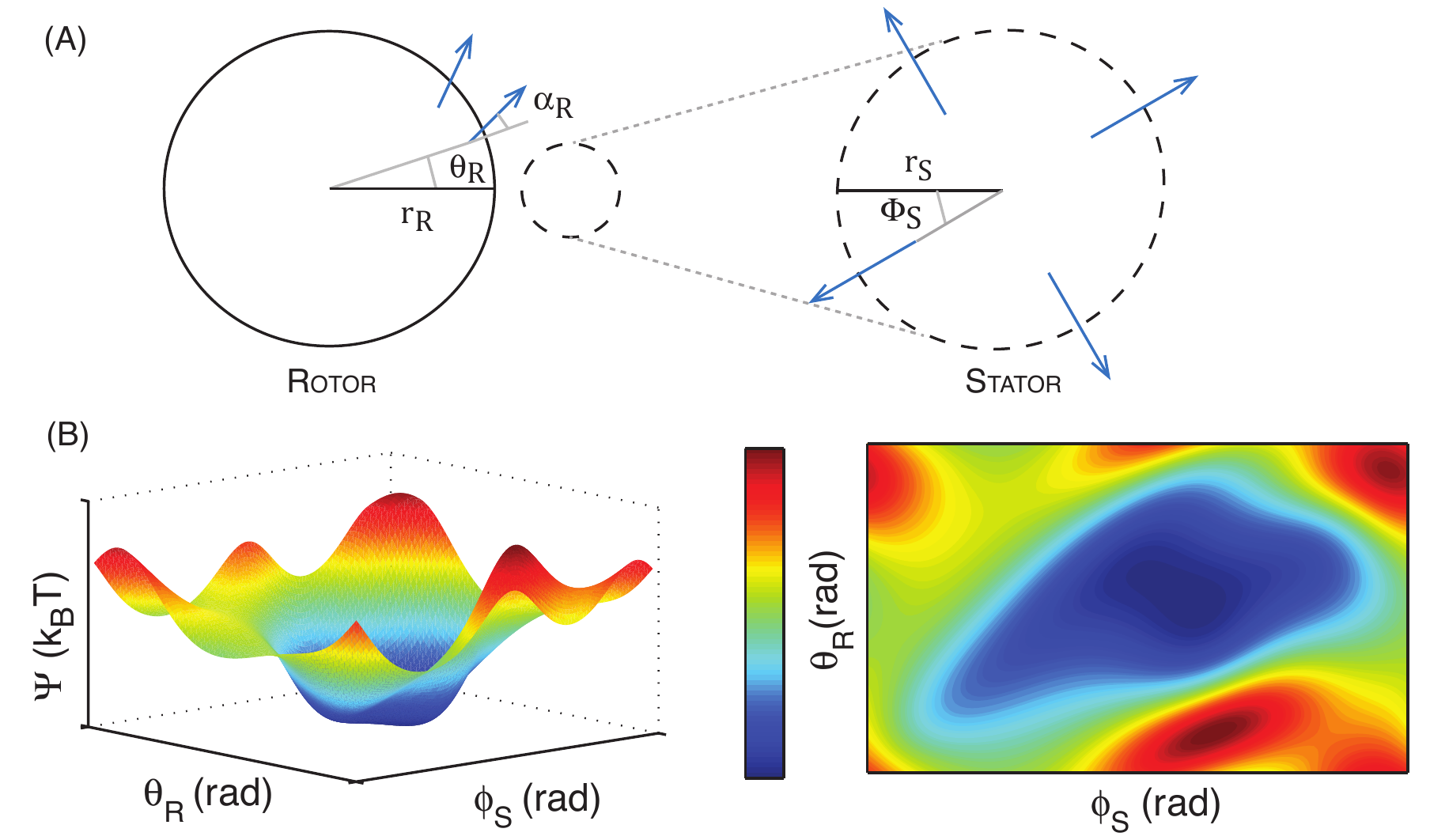}}
\caption{The predicted energy landscape during electrostatic steering. \textbf{(A)} Schematic of rotor and stator configurations. $\phi_S$ and $\theta_R$ are the angular coordinates of the stator and the rotor with respect to the horizontal; $\alpha_R$ is the positive angle of the individual FliGs with respect to the radius. Blue arrows denote the direction of the dipole \cite{Lee2010, brown2002crystal}. \textbf{(B)} Predicted surface and contour plots of the electrostatic energy vs. the stator and rotor angles. The predicted surface shows the existence of a wide and gently sloping energy well. Note that $\phi_S$ and $\theta_R$ are periodic variables with periods $\pi/2$ and $\pi/13$, respectively; the above plots show one period of each. Our calculations consider a single stator centered at (21,-2,1) with the rotor centered at the origin (all distances in nm). Computations using this dipole approximation suggest a well of depth $\sim$1 k$_\text{B}$T for this configuration (see the supplementary text for details).}\label{fig:electrostatic}
\end{center}
\end{figure}
The first step in our model is the steering and positioning of the stator by electrostatic forces. This hypothesis originates from the results of the swarming experiments performed by  Zhou~\textit{et~al.}~\cite{Zhou1995}. These studies were aimed at elucidating the structure of the MotA loops. They found that mutations of certain charged residues on the cytoplasmic portions of the loops hindered---but did not eliminate---motor function. Notably, the deleterious effect of mutations on the stator were often able to be countered by corresponding mutations on the FliGs. Certain mutations were also found to have very small effects, or even slight improvements, in bacterial motility. 

These results correspond to the idea that mutations of charged residues may result in imperfect steering and consequently in a less efficient, but still functioning, power stroke. Likewise, certain mutations may position the cytoplasmic loops closer to the adjacent FliG, resulting in a larger power stroke and thus improved motility.

Note that attractive electrostatic forces strong enough to comprise the entire power stroke would require a non-negligible energy to separate the stator and the rotor at the end of the power stroke. This penalty for `letting go' would obviate the rotor torque, resulting in a motor with a far lower Stokes efficiency \cite{wang2002stokes} than has been calculated for the BFM ($\sim$95\%) \cite{Berg2003}. In contrast, the mechanism we propose here efficiently generates mechanical work from the ion-motive force. 

Because detailed structural information of the stator is not yet available, we performed a simple example calculation to demonstrate how electrostatic interactions can serve to position the stator in place. Details on these computations can be found in the supplementary text. For computational convenience, we approximate the important charge residues on the FliG proteins \cite{brown2002crystal} and stator loops \cite{Zhou1998} implicated in torque generation as dipoles. The assumption that FliG proteins can be modeled as dipoles is based on previous studies \cite{Lee2010, brown2002crystal}. Modeling the electrostatic forces between the stator and rotor by point charge interactions produces results comparable to those obtained from a dipole approximation. 

Our calculations predict that the energy well produced by the electrostatic interactions will be shallow and wide (Figure~\ref{fig:electrostatic}). This results in a weak electrostatic force that is sufficient to position the MotA loop without significantly wasting energy to free the stator at the end of the power stroke. Furthermore, the width of the well leads to somewhat imprecise positioning. The distribution of observed rotor step-sizes has been shown experimentally to be centered around $\frac{2\pi}{26}$ radians ($\sim$13.8$^{\circ}$), the average spacing between consecutive FliGs \cite{sowa2005direct,nakamura2010evidence}. While this result is hardly unexpected, the wide spread of this distribution---in particular, the tendency towards smaller step sizes---has been somewhat puzzling. 

Because a wide energy well may result in stators being positioned at `nonoptimal' locations, electrostatic positioning may contribute to this variance. Since we propose that the stator's power stroke is imparted via a contact force on the rotor, imperfect electrostatic positioning will result in the stator being in contact for only a portion of its trajectory. This results in the stator delivering a stroke that is smaller than average. Of course, imperfect steering is not likely to be the only factor determining the variance in the observed step-size distribution: the uneven spacing of FliG's along the periphery of the rotor \cite{Paul2011}, as well as experimental errors, are also likely to contribute.

\subsection*{Motion about a proline hinge provides a steric push}
\begin{figure}
\centerline{\includegraphics[width=17.8cm]{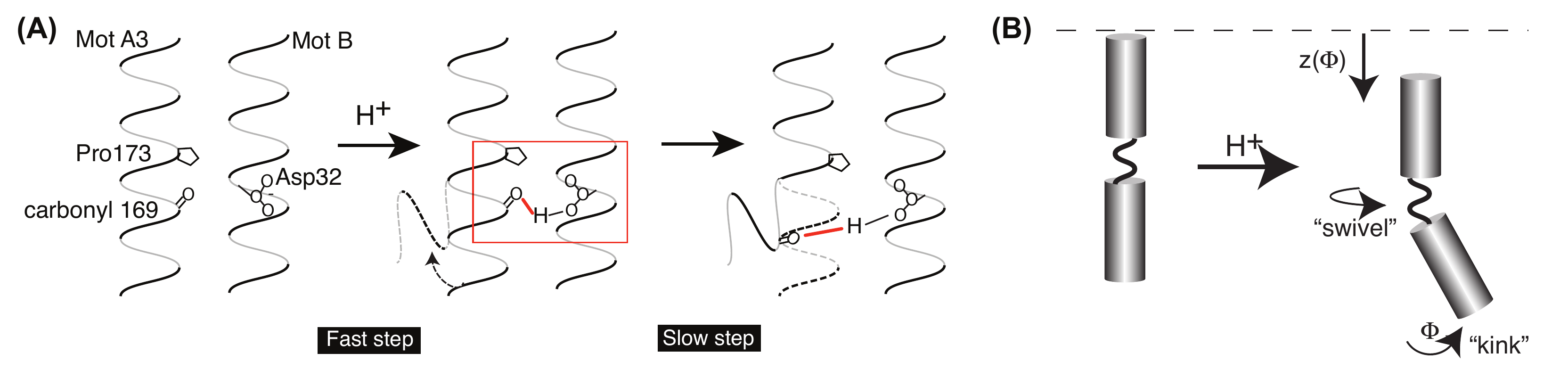}} 
\caption{Ion binding onto Asp32 induces a `kink and swivel' conformational change  \cite{Cordes2002}. 
\textbf{(A)} Binding of a proton to Asp32 of MotB drives a rapid local reorganization of hydrogen bonds (including those of water). In particular, we focus on the creation of a hydrogen bond between the side chain of MotB's Asp32 and the carbonyl group of MotA's residue 169. Ion binding thus creates a local elastic strain in the MotA helix. The release of this strain leads to the proposed conformational change in MotA about the Pro173 residue. Adapted from Kim \textit{et al.} \cite{Kim2008}.
\textbf{(B)} Upon ion binding, MotA undergoes a rapid conformational change consisting of three motions: (i) a bending about Pro173 $\phi$, (ii) a downward motion, $z(\phi)$, and (iii) a rotation about its central axis. Inspired by the work of Cordes \textit{et al.} \cite{Cordes2002}, we propose that this `kink and swivel' motion generates the power stroke. Importantly, we note that this figure is a two-dimensional depiction of a three-dimensional process, with the motion of the loop extending out of the plane of the page.}
\label{kns}
\end{figure}
As proposed previously \cite{Kim2008}, we hold that the steric portion of the power stroke is the result of a conformational change in the cytoplasmic MotA loop. In our model, this motion consists of hinged movements of the MotA helices that results in a movement termed `kink and swivel', as shown in Figure~\ref{kns} \cite{Cordes2002}. The steric mechanism proposed below remains valid regardless of which residue, or group of residues, on the MotA/MotB helices acts as the inducer. However, we have chosen to focus on MotA's Pro173 residue because: (i) along with Asp32 on MotB, this amino acid is strongly conserved across bacterial species \cite{Braun1999}, and (ii) previous molecular dynamics simulations have found that proline residues induce hinges in transmembrane helices~\cite{Cordes2002}, resulting in a movement analogous  to the one proposed in the model. The specific mechanism we propose is as follows. 

When a cation binds to the negatively charged Asp32 residue on MotB, the hydrogen bonds (including those of water) in the vicinity of Asp32 and Pro173 on the A3 helix of MotA collectively rearrange. This rearrangement induces an elastic strain in the MotA-MotB complex centered around the proline residue in the A3 loop of MotA. Figure~\ref{kns}A shows a candidate scenario, where the carbonyl group of residue 169 on MotA forms a hydrogen bond with Asp32 on MotB after proton binding, as proposed in  \cite{Kim2008}. This elastic strain induces the kink and swivel movement around the proline residue and drives the proposed motion of the lower part of the A3 helix, constituting the power stroke (see Figure~\ref{kns}B). The binding of the ion and the rearrangement of the hydrogen bonds (10$^{-12}$ to 10$^{-9}$ s) are near-instantaneous processes as compared to the much slower motion of the kink and swivel conformational change (10$^{-4}$ s). Thus the chemical steps can be treated as transitions between states in a Markov chain.

The above proposal is supported by a few simple calculations. The maximum torque of the BFM in \textit{E. coli} is $\sim$2000 pN-nm \cite{berry1997absence}. Given that up to 11 torque-generating units may be acting, this corresponds to a maximum motor torque of $\sim$200 pN-nm per stator \cite{Reid2006}. As the radius of the motor is $\sim$20 nm, the force generated by a single stator during a power stroke is $\sim$10 pN. Direct observation of stepping behavior have shown that the motor takes 26 elementary steps per revolution, corresponding to a displacement of $\sim$5 nm per step. As explained below, our model supposes that each elementary step is actually composed of two half-steps, each imparted by the power stroke of a MotA helix. This results in a displacement of $\sim$2.5 nm per power stroke. Molecular dynamics studies show the angles subtended by proline hinge motifs from various transmembrane helices to be between 18-25$^\circ$ \cite{Cordes2002}. From this, we can estimate the length of the cytoplasmic loop measured from the proline hinge to its tip to be $\sim$7 nm, a reasonable estimate as the majority of the stator residues have been shown to extend into the cytoplasm \cite{Zhou1995}. Such a lever arm would result in $\sim$25 pN-nm ($\sim$6-8 k$_\text{B}$T) of work per half-step, corresponding to the rearrangement of 1-2 hydrogen bonds. This energy barrier is sufficient to ensure an efficient directional process, as suggested in \cite{feng2008length}.

\subsection*{An in-phase two-cylinder engine}
\begin{figure}
\begin{center}
\centerline{\includegraphics[width=16cm]{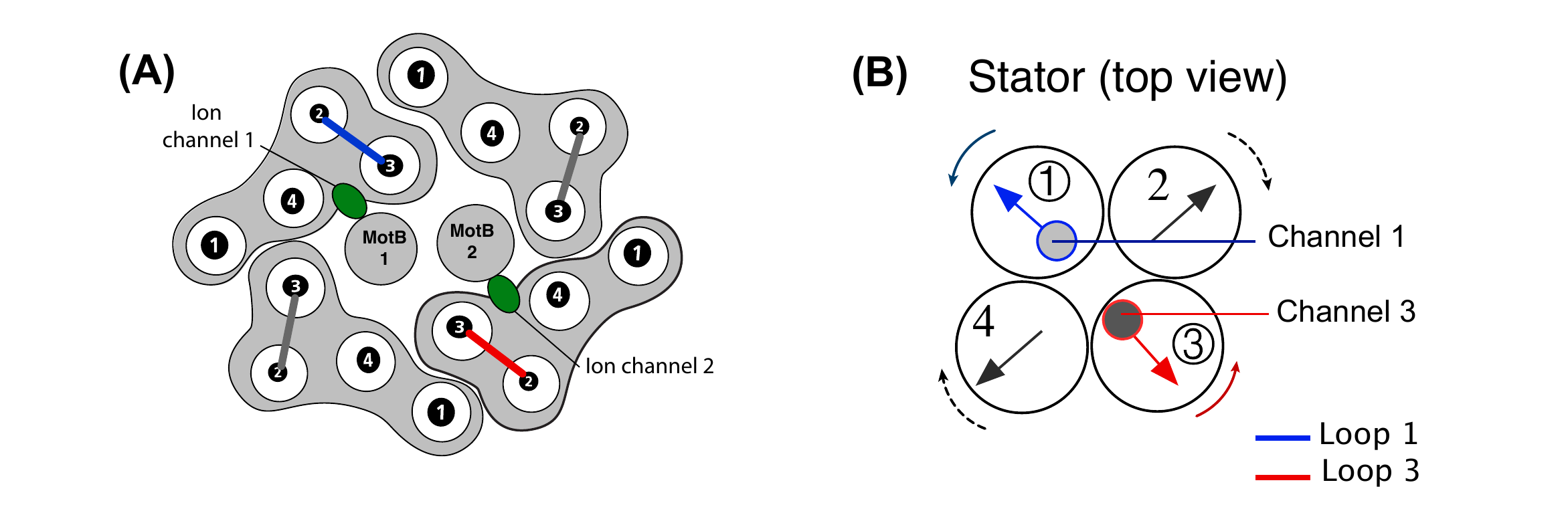}}
\caption{Stator structure and coordinated motion between stator subunits. \textbf{(A)} Proposed arrangement of stator components as viewed from the periplasm. A stator has four MotA helix bundles, each consisting of four $\alpha$-helices. The four MotA subunits surround a pair of MotB helices. The ion channels associated with the MotB's (shown in green) contain the Asp32 residue essential for proton binding. The stator is attached to the peptidoglycan via a linker region on MotB. The power stroke is delivered to the rotor FliGs by the cytoplasmic loops between helices A2 and A3 in each MotA bundle (shown as solid bars). Loops 1 and 3 (highlighted in blue and red, respectively) are associated with the ion channels. Adapted from Braun \textit{et al.} and Kim \textit{et al.} \cite{braun2004arrangement,Kim2008}. \textbf{(B)} Due to the helical structure of the MotA loops, we can make an analogy between their motion to that of a bundle of four gears. Our model proposes that Loops 1 and 3 (shown in blue and red, respectively) drive CCW rotation via contact with FliG, while Loops 2 and 4 drive CW rotation.}
\label{fig:structure} 
\end{center}
\end{figure}
There are four MotA subunits in each stator complex; see Figure~\ref{fig:structure} for a schematic of the stator structure. Our model supposes that two of these subunits are inactive during torque-generation while the motor is moving predominantly in a single direction. We base this presumption on the idea that switches between counterclockwise (CCW) and clockwise (CW) rotation  result from changes in FliG orientation \cite{Lee2010}. Given this, we propose that two MotA loops are responsible for the power stroke in one direction, while the other two interact with the alternately oriented FliG to drive rotation in the other direction. We suppose that loops 1 and 3 are responsible for CCW motion and loops 2 and 4 for CW motion, but note that this designation is arbitrary. This mechanism predicts that the intrinsic mechanics for power strokes in both directions are equivalent; this has been observed experimentally \cite{nakamura2010evidence}.

We propose that an elementary step is composed of a pair of power strokes, analogous to the mechanism of a two-cylinder engine. It may be possible to observe these substeps in experiments on motors driven at extremely low speeds. This can be done using chimeric sodium-driven flagellar motors. As extremes in sodium concentration are tolerated far more easily than extremes in pH, these chimeric motors can be driven at very low sodium motive forces (SMFs). Thus far, speeds as low as 10 Hz have been obtained \cite{Sowa2005}. 

A two-ion mechanism  can either be `in-phase', in which the energetic profiles of the two stator loops are identical, or `out-of-phase', in which their dynamics are offset by a half-cycle. In an experiment using a slowly-driven chimeric motor, measuring the rate-limiting step between mechanical substeps can differentiate between these two scenarios. For example, if slower ion-binding (e.g., by lowering IMF) extends the dwell time between half-steps, the out-of-phase engine model is supported. 

The mechanics of these two scenarios are equivalent within the framework of our model. For this reason, we discuss only one of these mechanisms in detail: the one in which the two stator loops act in-phase with each other (as shown in Figure~\ref{fig:mechanism}B). We choose this alternative because the passage of a single proton across a membrane is unlikely to be sufficient to generate an efficient power stroke. A proton passage under standard conditions generates $\sim$6 k$_{\text{B}}$T, which is slightly less than the calculated length of `time's arrow' (the energy barrier required for a process to be reliably directional in the presence of thermal noise) \cite{feng2008length}.

\subsection*{A full revolution requires the passing of at least 52 protons}

Our model for torque generation assumes that the rotation of the BFM is `tightly coupled' to the transmembrane ion gradient. This means that each elementary power stroke is tied directly to the passage of protons across the membrane. Given our prior assumption of 26 elementary steps per revolution, our model thus requires 52 protons for a full revolution. Previously, a lower bound for the number of ions per full revolution was determined by calculating the work done as $\langle\tau\rangle \times 2\pi$ and equating it to the free energy $n \times$IMF, where $n$ is the number of ions per revolution and IMF is the ion motive force, as before \cite{lo2013mechanism}. The above calculation resulted in an estimate of $n=37$, lower than the 52 ions per revolution supposed by our model.

This discrepancy can be explained as follows. While the above is indeed a lower bound, a tighter bound can be computed. The calculation of work as stated above suggests that the power output per revolution is $\widetilde{P} = \langle\tau\rangle \langle\omega\rangle$. However, power is formally calculated as $P = \langle\tau \cdot \omega\rangle$, which differs from $\widetilde{P}$ by a covariance term, $\text{cov}\left(\tau,\omega\right)$. This follows from the fact that, for any two stochastic processes $X$ and $Y$, $\langle XY\rangle = \langle X\rangle \langle Y \rangle + \text{cov}(X,Y)$.

Note that the number of protons per revolution assumed by our model is also a lower bound; that is, we have assumed that 52 \textit{working} ions are required per revolution. Many factors can result in the passing of more ions than predicted, including leakiness of the ion channels, irregular arrangement of FliG's around the rotor, or imperfect placement of stators by the electrostatic steering forces. This can be quite easily extended within our mathematical framework by replacing the step function associated with ion binding with a sigmoidal function.

\subsection*{The mechanical escapement} 

\begin{figure}[h!]
\begin{center}
\centerline{\includegraphics[width=17cm]{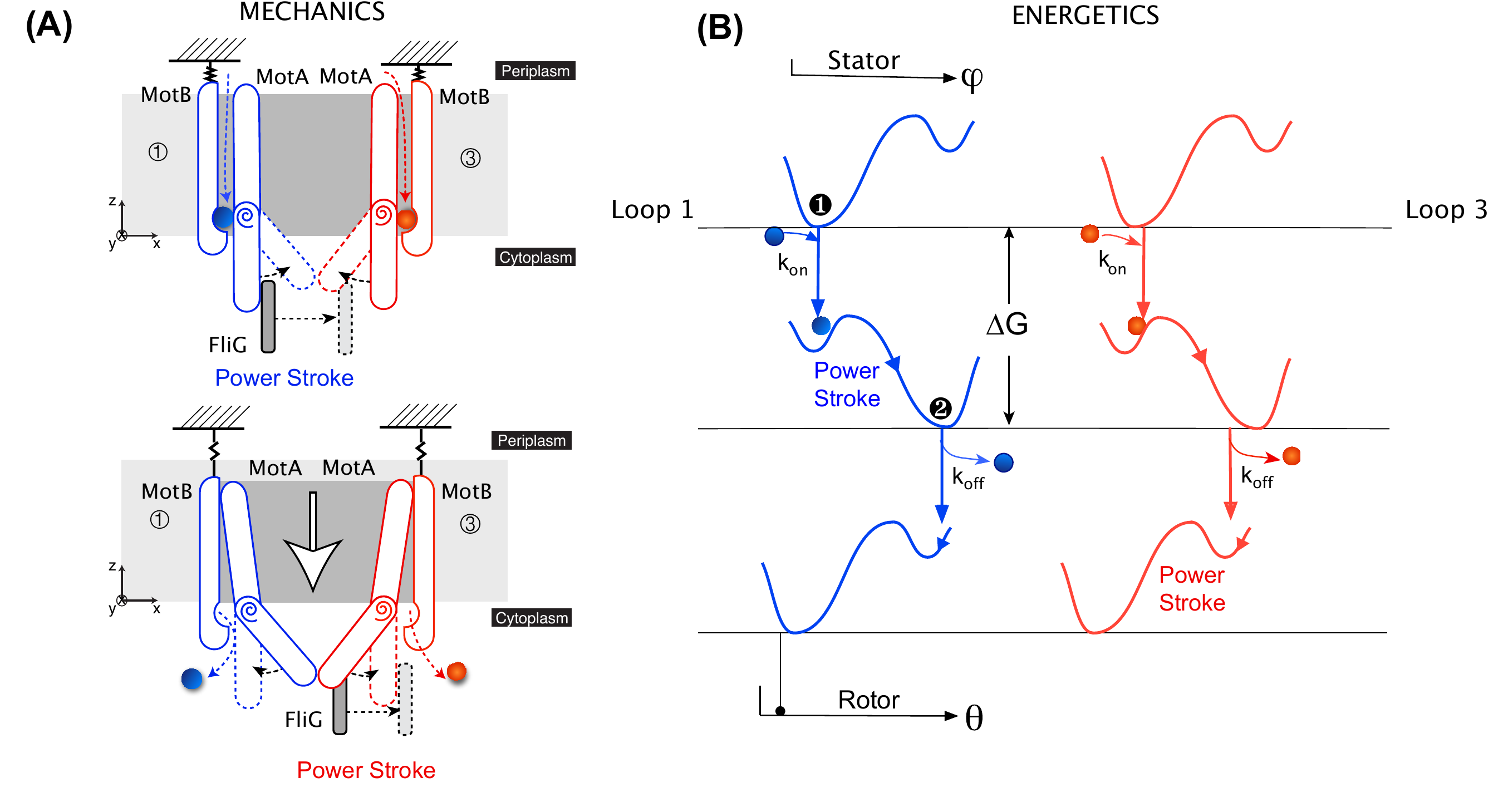}}
\caption{Dynamics of the rotor-stator interaction. \textbf{(A)} \textsc{Mechanics of the power stroke}. \textit{Top panel:} After the initial electrostatic steering, two protons bind to the charged Asp32 residues on the MotB's. The consequent rearrangement of hydrogen bonds induces an elastic strain in the straight MotA loops. Release of this strain results in synchronous `kink and swivel' motions about the proline `hinge' in both MotA's. As a result, a steric push is imposed on FliG, and the first half of the power stroke performed by Loop 1. Importantly, this motion also has a vertical component---the loops lower themselves out of the membrane. \textit{Bottom panel:} The lowering of the MotA loops exposes the protons in MotB to the cytoplasm, where they are released. This results in a `reset' of the MotA loops, during which the Loop 3 carries out the second half of the power stroke. We note that this image depicts a two-dimensional projection of a three-dimensional motion: the motion of the stators is not constrained to the plane of the page. \textbf{(B)} \textsc{Energetics of the power stroke}. Because the two loops move in-phase with each other in our model, their energetic pictures are identical. We describe the free energy landscapes using double-well Landau potentials. These landscapes are shown in blue for Loop 1 and red for Loop 3 with respect to the angles of the stator $\phi$, and rotor $\theta$.  The initial entrance of the proton into the ion channel ($k_{\text{on}}$) places the system within $k_BT$ of the energy barrier. Thermal motions then result in the first half of the power stroke (\textit{Top and middle panels}). The exit of the protons into the cytoplasm ($k_{\text{off}}$) results in the `reset', and the second half of the power stroke (\textit{Middle and bottom panels}).}\label{fig:mechanism}
\end{center}
\end{figure}

Figure \ref{fig:mechanism}A depicts the mechanics associated with the power-stroke. We choose the angle subtended by a stator loop $\phi_S^i$ (where $i$ corresponds to the loop number) as the order parameter. That is, we consider the energy landscape along the arc length of the mechanical trajectory of the stator loop. A stator loop has two stable configurations: straight ($\phi_S^i = 0^{\circ})$ and bent ($\phi_S^i  \sim 20^{\circ})$. Both of these configurations correspond to energy minima in different chemical environments: when the negative Asp32 is not neutralized by a proton, the loops prefer to maintain a straight posture ($\phi_S^i  = 0^{\circ})$. The presence of bound protons induce a free energy change sufficient such that a thermal fluctuation can induce the conformational change to the bent state ($\phi_S^i  \sim 20^{\circ})$. 
During a power stroke, the entire stator complex undergoes a collective gear-like motion as shown in Figure~\ref{fig:structure}B. The conformational change due to the `hopping on' of the ion produces the first half of the power stroke: here, loop 1 pushes the FliG, while loop 3 is put in place to carry out the second half of the power stroke during the reset (Figure~\ref{fig:mechanism}A). This `reset' corresponds to the `hopping off' of the proton, resulting once again in the the stator loops surmounting the energy barrier between configurations and reverting to the straight position ($\phi_{1,3} = 0^{\circ}$). Note that the numbering of the loops is  arbitrary; the mechanism proposed here is equivalent to one in which loop 3 performs the first half of the power stroke and loop 1 the second. 

In summary, a torque-generation cycle by a single stator of the bacterial flagellar motor proceeds as follows:
\begin{enumerate}
\itemsep0em
\item Electrostatic interactions between charged residues on MotA and FliG steer a stator tip close to a rotor FliG.
\item In the presence of a membrane potential, the two MotB aqueous ion channels open and two protons bind to the negatively charged Asp32 residues on the MotBs. This triggers a reorganization of the hydrogen bonds in the vicinity of the Pro173 on MotA (see Figure \ref{kns}A). 
\item The hydrogen bond rearrangements induce elastic strain in the straight MotA loops. This strain drives a `kink and swivel' motion of the MotA loop, increasing the bend angle (from $\phi_S^i  = 0^{\circ}$ to $20^{\circ} $, as shown in Figure \ref{kns}B).
\item One MotA loop (Loop 1, shown in blue in Figure \ref{fig:mechanism}A) applies a steric push to the nearest FliG, resulting in one-half power stroke.
\item At the same time, the movement of the stator ion binding pocket moves downward so that the pocket is exposed to the cytoplasm. The ion-channel is now closed to the periplasm. The protons `hop off' MotB into the cytoplasm, now inverting the strain in the bent MotA loops .
\item The inverse strain drives the movement of the loops in the reverse direction, straightening the bent MotA's (i.e., from $\phi_S^i  \sim 20^{\circ}$ to $0^{\circ}$).
\item The other MotA loop (Loop 3, shown in red in Figure \ref{fig:mechanism}A) now applies a steric push to the same FliG, completing the second half of the power stroke. 
\end{enumerate}
Consequently, to the rotor, the stator appears to be an `inch-worm'  stepper with FliGs as the `stepping-stones'.

\begin{figure}
\begin{center}
\centerline{\includegraphics[width=17cm]{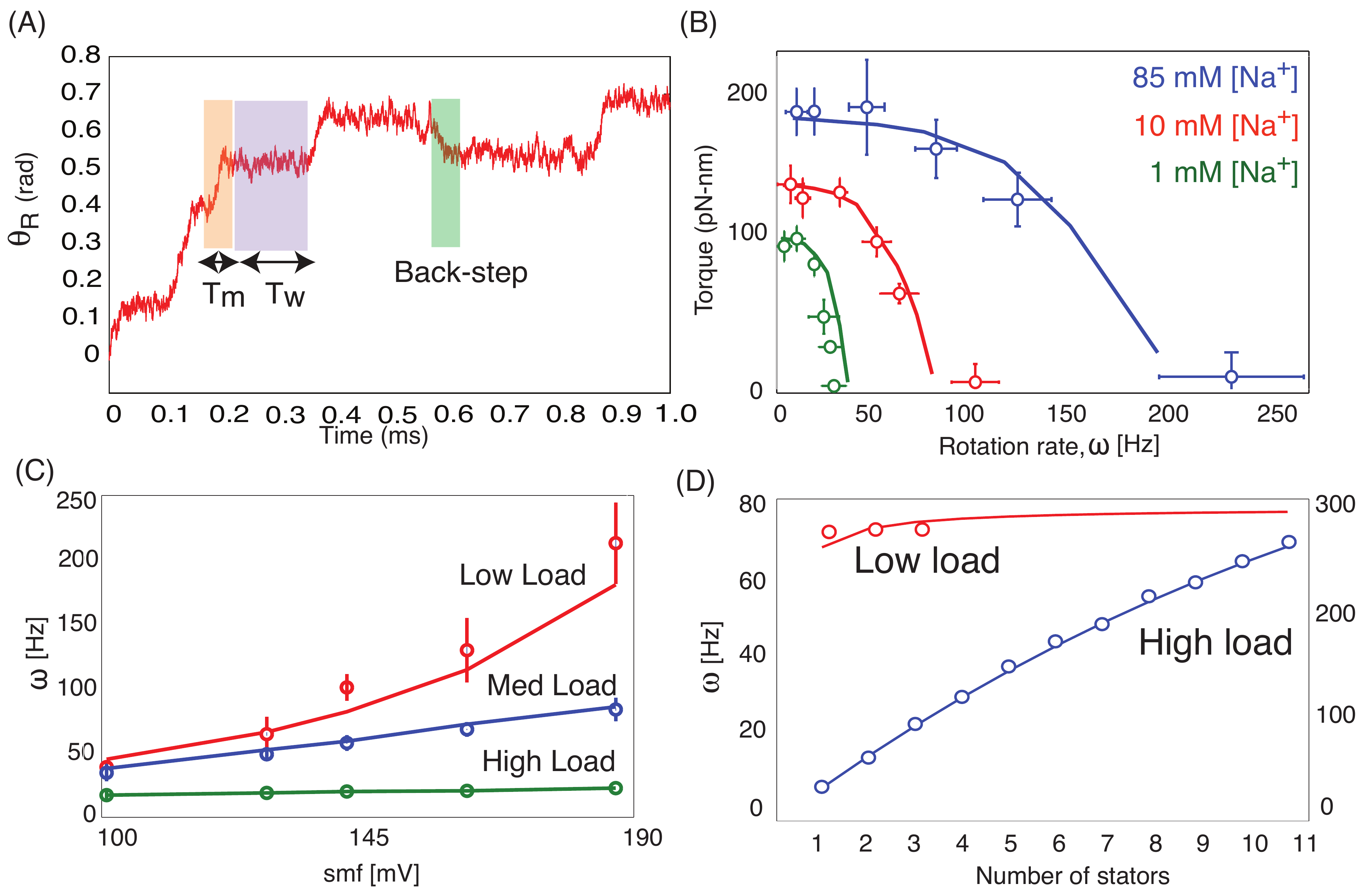}}
\caption{Summary of recent experiments and comparisons with simulations of our model. Results in \textbf{(A,B,C)} are derived from numerical simulations. Results in \textbf{(D)} are analytic curves obtained from a deterministic approximation (see the supplementary text). In all plots, model calculations are shown by solid lines, and experimental data is shown as open colored circles. \textbf{(A)} Sample trajectory generated by the model. Moving ($T_M$) and waiting ($T_W$) times are shown with orange and purple backgrounds, respectively. Occasionally, reversals (shown with green background) appear when MotA Loops 2 and 4 are engaged due to conformational changes in FliG. \textbf{(B)} Single-stator torque-speed curves measured in a chimeric sodium motor for various sodium concentrations at pH 7.0. Curves show a concave-down shape, with the length of their 'plateau' being SMF-dependent (data from Lo \textit{et al.} \cite{lo2013mechanism}). \textbf{(C)} Motor speed vs. SMF in a chimeric sodium motor shows a nearly linear relationship across various loads (data from Lo \textit{et al.} \cite{lo2013mechanism}). \textbf{(D)} The maximum speed at low and high load may show different relationships with the number of stators (data from Sowa and Berry  \cite{sowa2008bacterial}).  We note that this result is dependent on the assumption that all stators work in synchrony, and that it has come into question given recent results from the Berg laboratory \cite{Lele2013}.}
\label{fig:data}
\end{center}
\end{figure}
\section*{Results and Predictions}

Using the mathematical model described in the Materials and Methods, we performed both analytic calculations and numeric stochastic simulations. Statistics from simulated trajectories---an example of which is shown in Figure~\ref{fig:data}A---were used to calculate various experimental quantities including average motor torque and angular speed. 

In the sample trajectory for the rotor motion, the duration of a power stroke ($T_m$) and the waiting time between consecutive power strokes ($T_w$) are highlighted in orange and purple, respectively. The highlighted power stroke shows two half-steps, corresponding to the two sequential steric pushes by the two MotA loops involved. As in experimental trajectories, occasional reverse steps are also observed in our simulations, one of which is shown in the sample path in Figure~\ref{fig:data}A. An explanation for back-steps that is compatible with our model is provided below.

The results shown in Figure \ref{fig:data}A,B,C were obtained via simulation. Analytic calculations on an approximate deterministic model (explicitly provided in the supplementary text) were also performed. The results shown in Figure \ref{fig:data}D were computed analytically. Finally, in addition to fitting current experimental data, we propose further experiments to test, and hopefully validate, several of our model predictions (e.g., Figure \ref{fig:pred}). 

\subsection*{Single-stator motors exhibit concave-down torque-speed curves}

Until recently, BFM experiments have been performed on motors with multiple stators, with no direct accounting for the number of engaged stators at a given load. Therefore, the existence of the torque-speed `plateau' and `knee' have been assumed to be innate characteristics of the rotor-stator interaction, largely because there was no evidence to the contrary. However, Lo \emph{\textit{et al}.} \cite{lo2013mechanism} performed experiments using a chimeric single-stator motor showing smoother torque-speed curves without a dramatic plateau as observed for wild type motors. While these curves are still concave-down in shape, the extent of the plateau regions are quite variable, and are dependent on the IMF. 

Our simulations show torque-speed relationships consistent with these single-stator experiments (Figure~\ref{fig:data}B). The behavior of the torque-speed curves results from a competition between the time taken for a mechanical half-step ($T_M$) and the waiting time between ion-binding events ($T_W$). For example, our simulations show that the average time in moving a half-step $\langle T_M \rangle$ is $\sim$2 to 20 ms at high loads and $\sim$0.01 ms at low loads. The average waiting time under standard conditions $\langle T_W \rangle$ is $\sim$0.2 ms \cite{meacci2009dynamics}. Therefore, at low loads, the motor is in a kinetically limited regime, where the waiting time between steps is generally higher than the time required to complete a step. Conversely, the motor is mechanically limited at high loads when $\langle T_M \rangle > \langle T_W \rangle$, resulting in the observed plateau. Consequently, as shown in Figure~\ref{fig:data}B, this plateau region grows smaller as the IMF decreases (i.e., as $\langle T_W \rangle$ increases).

This competition is also manifested in the relationship between speed and IMF: speed depends linearly on IMF at high loads, but in a slightly nonlinear fashion at low loads (Figure~\ref{fig:data}C). Given that the rotor moves $2\pi/26$ radians per step, the speed of the rotor ($\omega_R$) can be approximated as:
\begin{equation*}
\omega_R  \approx \displaystyle \frac{2\pi}{26} \times \frac{1}{ \langle T_M \rangle +  \langle T_W \rangle}.
\end{equation*}
At high loads, $\langle T_M \rangle \gg \langle T_W \rangle$. Because the force applied during a power stroke is inversely proportional to the ion-motive force, $\langle \omega_R \rangle \propto$ IMF at high loads. In contrast, the waiting time eclipses the time for a mechanical step at low loads, and therefore $\omega_R \propto 1/\langle T_W \rangle \propto \exp(\text{IMF}/k_BT)$. Further details to this end are provided in the supplementary text. 

\subsection*{Backstepping in the absence of CheY-P is due to thermal flipping of FliG}

The BFM plays a central role in bacterial chemotaxis:  the direction of rotation of the motor determines whether a bacterium will move in a straight line (CCW) or `tumble' (CW) to move in a random new direction. This switching is typically initiated via a signal transduction pathway, in which a response regulator protein, CheY, is phosphorylated into an activated form, CheY-P, to induce tumbling. For more information on this pathway and bacterial chemotaxis, we refer the reader to several excellent reviews \cite{wadhams2004making,porter2011signal}.

However, occasional backsteps (e.g., CW motion during primarily CCW rotation) are observed even in the absence of CheY-P. This has been attributed to microscopic reversibility, of which three possible models are discussed in the supplementary text. For example, Mora \textit{et al.} ascribed switching in the BFM to the diffusive motion of the rotor through a `bumpy' 26-fold periodic potential \cite{mora2009steps}. However, recent structural studies have found that there exist two main configurations for the FliGs \cite{Lee2010,nakamura2010evidence}, lending support to the idea that a `flipping' between these states is the molecular basis for backstepping. We note that despite a general agreement on the \textit{existence} of two distinct FliG configurations, the exact nature of the conformational change to the CCW direction remains controversial.

In our model, the probability of observing a backward step is equivalent to the probability of finding a FliG oriented in the CW state (assuming a primarily CCW-rotating motor). Within the framework of our model, whenever a FliG changes its state and is close to a stator, then the stator uses loops 2 and 4 to apply a contact force and pushes the FliG in the CW direction. To model the flipping between CW and CCW states for the FliGs, we use a nearest-neighbor periodic Ising model with the 26 FliGs arranged on a one-dimensional ring. Such models have been used successfully to explain rotational switching (see, e.g., \cite{Duke2001}). 

In our model, when the FliGs are oriented at an angle of roughly 10-20$^{\circ}$ with respect to the radial direction, as shown in Figure \ref{fig:electrostatic}A, the motor moves in the CW direction by virtue of contact forces from loops 2 and 4. Conversely, when the FliGs are pointed either orthogonal or at an angle of 180$^{\circ}$ with respect to the CW orientation, the motor steps in the CCW direction using loops 1 and 3. Using an Ising model for the flipping of FliGs, we calculate the probability of a backstep to be $\sim$8\%. This probability was calculated to be $\sim$7.3\% from stepping statistics collected by Sowa \emph{et al.} \cite{sowa2005direct}, demonstrating that a backstep might indeed be simply due to fluctuations in FliG orientation. Further details on these calculations are provided in supplementary text~S1. 

\subsection*{Speed and stator number are nonlinearly related at both high and low loads}

Recent experiments have shown that the bacterial flagellar motor is a dynamic structure, constantly exchanging stators in response to changing viscous loads \cite{Lele2013,leake2006stoichiometry,tipping2013load}. Recently Tipping \emph{et al.} \cite{tipping2013load} showed that the number of torque-generating stator units depends on the viscous load. Dynamic studies have suggested that new stators may come online as the load increases \cite{Lele2013}. This result appears to be responsible for the `resurrection' experiments which show a stepwise increase in speed as stator units are recruited at high loads, but only a single jump at low loads \cite{sowa2008bacterial}. 

Under the assumption that individual stator units act in synchrony with each other, the mathematical model provided in the Materials and Methods can be extended to deal with the dynamics of motors with multiple stators. The details of this extension and generalized equations for motors with $N$ stators are provided in the supplementary text. The speed shows a slightly nonlinear dependence on the number of stators at very high loads (Figure~\ref{fig:data}D). A linear relationship between speed and stator number would imply that the applied force of the stator is independent of the load; our results suggest that this nonlinearity may arise as a natural consequence of the steric force.

Furthermore, our computations match results from the resurrection experiments very well, even when a constant number of stators is assumed (Figure~\ref{fig:data}D). Given the assumptions of the extended model, we predict that the motor speed at low loads will be independent of stator number if (and likely, only if) individual stators step in synchrony. Further experiments are needed to determine if this is indeed the case. Because of the recent experiments performed on single-stator motors, we have chosen to primarily focus on the innate mechanism of torque-generation in the BFM. For this reason, a more thorough investigation of the interaction between stators, while worthwhile, is left for a future study.
\begin{figure}
\begin{center}
\centerline{\includegraphics[width=16cm]{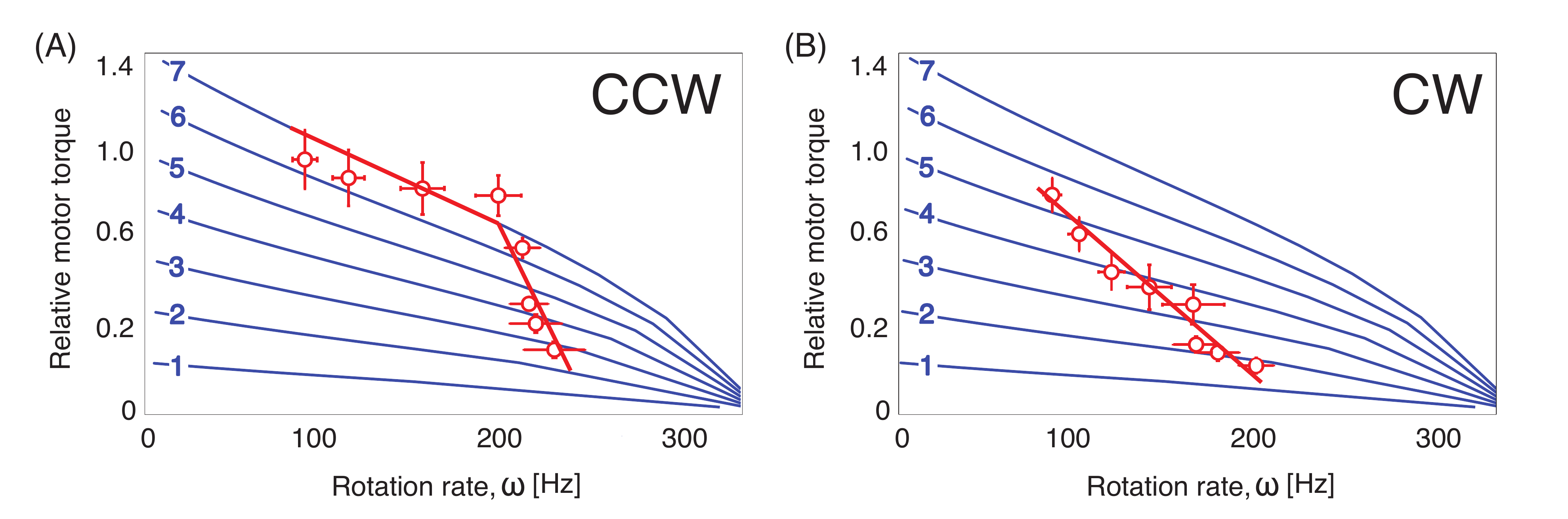}}
\caption{Prediction of how the number of activated stators at a given load and torque-speed relationship in a constant-stator motor can reproduce observed torque-speed curves. Red open circles show data from Yuan and Berg \cite{yuan2010asymmetry} for motors movithe supplementary textng in the \textbf{(A)} counter-clockwise (CCW) and \textbf{(B)} clockwise (CW) directions. Data was collected from motors in which the number of engaged stators was not constant across loads. Analytic curves from our model for motors with $N=1,\dots,7$ stators are shown in blue. The asymmetry between the observed CCW and CW torque-speed relationship is likely due to differences in stator recruitment \cite{tipping2013load}.}
\label{fig:pred}
\end{center}
\end{figure}
\subsection*{The observed constant-torque plateau is due to stator recruitment}

Dynamic experiments by Lele \textit{et al.} showed that the number of active torque-generating units in the BFM is not constant over all applied loads \cite{Lele2013}, thus bringing into question current  interpretations of the well-known torque-speed curve \cite{yuan2010asymmetry}. Our model simulations are consistent with the single-stator curves measured by Lo \textit{et al.}, lacking the large constant-torque plateau seen in previous experiments where the number of stators was not directly accounted for. 

Given this, we suggest that the observed two-regime curve is a consequence of variable stator number rather than the rotor-stator interaction. Torque-speed curves predicted by our extended model for varying number of stators are shown as solid blue lines for $N=1,2,\dots,7$ stators, overlaid with experimental data (red circles) for CCW  (Figure \ref{fig:pred}A) and CW (Figure \ref{fig:pred}B) motors. The relative motor torque is shown as in the experiment: normalization is performed relative to the torque generated by a motor with 6 stators moving at $\sim$70 Hz (the lowest speed measured in \cite{yuan2010asymmetry} for the CCW motor). 

In particular, as shown in Figure \ref{fig:pred}A, we predict that the `constant torque plateau' seen in CCW motors is due to the motor maintaining a constant number of stators, and the sharp transition to decreasing torque is due to stators becoming inactive at a critical load or speed. Furthermore, as shown in red in Figure \ref{fig:pred}B, torque-speed measurements on CW motors yield a near-linear torque-speed relationship \cite{yuan2010asymmetry}. In accordance with our prediction, this asymmetry was reflected in the experiments of Tipping \textit{et al.}: CCW motors cease to recruit additional stators at loads far lower than CW motors \cite{tipping2013load}. This suggests that CW motors `jump' between curves as they recruit additional stators at low loads, but stay on a single curve (i.e., maintain a constant number of stators) once the load is sufficiently high, while CW motors continually recruit stators at loads far lower than CCW motors. 

To this end, our results support the idea that the characteristic `knee' commonly associated with the dynamics of the flagellar motor may simply be an artifact of experiments performed on motors with a variable number of stator units \cite{Lele2013}, rather than an intrinsic feature of the rotor-stator interaction. This further asserts that a model for the loading and unloading of torque-generating units is needed to completely interpret the torque-speed curves for experiments where the number of active stators was not monitored.  

\section*{Discussion}

The ability to convert a transmembrane ion gradient into rotary torque is rare, observed in only two known protein motors: the F$_\text{O}$ motor of ATP synthase and the BFM. The mechanism behind the torque generation in the latter has been a longstanding mystery, driven by the fundamental role of this machine in bacterial locomotion and chemotaxis.

Recent experiments have shown that the number of torque-generating units in the flagellar motor is likely dynamic and not constant across previously measured torque-speed curves~\cite{Lele2013}. This result brings into question previous models of torque-generation that have relied on these curves. Here we combined recent biophysical \cite{lo2013mechanism} and structural \cite{Kim2008,Lee2010} studies to reexamine these models. 

Crucially, we have utilized all known structural information of the BFM, as well as the experimental measurements on single-stator motors by Lo \textit{et al.} \cite{lo2013mechanism}, to present the first mechanically specific model of torque generation. Using this information, we are able to present an explicit model of the dynamics of the stator during a torque-generation cycle. Our model implicates a steric interaction between the cytoplasmic MotA stator loops and the FliG proteins of the rotor. We have predicted that this interaction is driven by conformational changes in the MotA loops due to the binding of a cation to an essential aspartate residue on MotB. Results from our model simulations reproduce recently measured torque-speed and speed-IMF curves from single stator motors.

The model we have proposed is akin to a two-cylinder engine, where two of the four MotA loops act when the motor is moving in the counterclockwise direction and the other two loops act in the clockwise direction. We have proposed that the two loops act `in-phase' with each other, moving in synchrony as two protons bind to the MotBs and are subsequently released into the cytoplasm. In this manner, the first loop executes its half of the power stroke when the protons are bound to the MotBs, and the second does so once they have `hopped off' into the cytoplasm. 

Experiments performed at low IMF can be used to differentiate between a stator acting as an engine which is `in-phase' or 'out-of-phase'. Because ion-binding is rate limiting under these conditions, trajectories would show clear half-steps only if the BFM acts as an `out-of-phase' engine. However, given that the mechanics of the power stroke for both scenarios are equivalent, the mechanism corresponding to an out-of-phase engine would lead to a calculation analogous  to the one presented in this work.

A notable prediction of our model arises from extending our equations for single-stator motors to model the dynamics of motors with multiple stator torque-generating units. Our results predict that the observed two-regime torque-speed curve is likely the result of a load-dependent recruitment of stators (Figure \ref{fig:pred}A), with the sharp transition between regimes corresponding to the point at which stators begin to come off the motor. This lends strong support to the theory that the unique shape of the BFM torque-speed relationship may only exist due to the inability of most experiments (until now) to explicitly monitor the number of engaged stators at any given load \cite{Lele2013}. This further indicates that a model of stator recruitment is necessary to fully understand experimental results performed on motors where the number of stators was not necessarily constant. Furthermore, it suggests that low-load measurements were never successfully performed on motors with multiple stators engaged, thus shedding doubt on the result that increasing stator number does not effect motor speed at low loads.

\section*{Methods}
\begin{table}[t]
\caption{List of parameters used with units, values, and reference.}
\begin{center}
\begin{tabular}{ccccc}
Parameter & Definition  & Units & Values & Ref \\ \hline
$R$ & Radius of the rotor  & nm  & 20  &  \cite{Berg2003} \\
 $\ell_P$ & Length of the proline hinge arm  & nm  & 7  &  \cite{Zhou1995}\\
$\zeta_S$ & Drag coefficient of the stator  & pN-nm-s-rad$^{-1}$ & 0.0002 & - \\
$\zeta_R$  & Drag coefficient of the rotor  & pN-nm-s-rad$^{-1}$ & 0.017 & - \\
$\zeta_L$ & Drag coefficient of the load & pN-nm-s-rad$^{-1}$ & 0.0--10  & \cite{yuan2010asymmetry}\\
$\phi_S$ & Angular position of the stator & rad & - & - \\
$\theta_R$ & Angular position of the rotor & rad  & -  & - \\
$\theta_L$ & Angular position of the load & rad &  - & - \\
$\kappa$ & Hook spring constant & pN-nm-rad$^{-1}$ & 1000 & \cite{block1989compliance} \\
$N$ & Number of stators & - & 1--11 & \cite{yuan2010asymmetry} \\
$\tau$ & Rotor torque from stator & pN-nm & - & - \\
$\Psi$ & Electrostatic potential & - & 1.5--2 k$_\text{B}$T & - \\
\hline
\end{tabular}
\end{center}
\label{tab:params}
\end{table}
The mechanochemistry of the torque-generation cycle of a flagellar motor with a single stator unit can be modeled by the following Langevin equations. The dynamics of the angular positions of the stator loops $\phi^{i}_S(t),$ $i \in \{1,3\}$ are given as:
\begin{equation}
\small \zeta_S\frac{d\phi_S^i}{dt} = \underbrace{-\frac{\partial G}{\partial\phi_S^i}\ell_p}_{\substack{\text{Torque from}\\\text{Proline hinge}}} - \underbrace{\frac{\partial V_{RS}}{\partial \phi_S^i}}_{\substack{\text{Reaction}\\\text{from rotor}}} -  \underbrace{\frac{\partial\psi}{\partial\phi_S^i}\ell_p}_{\substack{\text{Electrostatic}\\\text{attraction}}} + \underbrace{\sqrt{2k_BT\zeta_S}f_n(t).}_\text{Thermal fluctuations}
\end{equation}
Here, as in the following equations, the last term is the stochastic Brownian force, where $N(t)$ is uncorrelated white noise. $\zeta_S$ is the effective drag coefficient of the stator. $G = G(\phi^{i}_S,j)$ denotes the free energy of stator loop $i$, modeled in Figure \ref{fig:mechanism}B as a Landau potential. However, because of thermal fluctuations the exact shape of the potentials is immaterial. Accordingly, we approximate this potential by a piecewise linear function for ease of computation. The parameter $j \in~\{0,2\}$ corresponds to the chemical state of the system: $j = 2$ if two protons are bound to the MotB helices and $j=0$ if not. The switching between the two chemical states corresponds to a `jump' between potential curves, as shown in Figure \ref{fig:mechanism}B. 

As the stator moves between the two configurations, it induces a contact force, and subsequent torque, on the rotor. Unlike previous models, we do not assume that this torque is constant across loads, but rather depends on the $\zeta_L$ (see the supplementary text for more information). To this end, we do not allow a linear interaction potential between the stator and the FliG; this would result in a constant applied force, which is not true for contact forces. We model the steric interaction potential $V_{RS}$ as
\[
V_{RS}(\phi_S^{i},\theta_R)  = 
 \begin{cases}
    -F_{RS}\frac{(R\theta_R - \ell_P\phi_S^i)^2}{X_{RS}} & \text{if } 0 \leq x \leq X_{RS} \\
   0       & \text{otherwise}, 
  \end{cases}
\] 
where $x = X_{RS} + R\theta_R - \ell_p \phi^i_S$ denotes the distance between the position of the stator loop and the nearest FliG. From this, the torque imposed on the rotor is calculated as $\tau_{\text{contact}} = -\partial V_{RS}/\partial\theta_R$, while the corresponding reaction torque on a stator loop is given by $ \tau_{\text{reaction}} = -\partial V_{RS}/\partial \phi_S^{i}$. 

The charges on the FliG and the stator loop exert weak attractive forces on each other. These forces prevent the drifting of the rotor with respect to the stator during the chemical transition events. We refer the reader to the supplementary text for more on the effects of attractive electrostatic forces on the torque-speed curves. With the contact torque and the weak electrostatic forces, the total instantaneous torque on the rotor is given by $\tau = \tau_{\text{contact}} - R(\partial\psi/\partial\theta_R)$. The average torque on the rotor is a time (or ensemble) average of the instantaneous torque. Finally, the rotor and load are connected by a linear spring with elastic constant $\kappa$; the elastic coupling terms in the equations for the rotor and the load thus appear with opposite signs. 

Given this, the rotor dynamics are described by a corresponding Langevin equation:
\begin{equation}
 \small \zeta_R\frac{d\theta_R}{dt} =  \underbrace{-\frac{\partial V_{RS}}{\partial\theta_R}}_{\substack{\text{Torque}\\\text{from stator}}} - \underbrace{\frac{\partial\psi}{\partial\theta_R}R}_{\substack{\text{E.S.}\\\text{attraction}}} - \underbrace{\kappa(\theta_R-\theta_L)}_{\substack{\text{Connection}\\\text{to load}}}+ \underbrace{\sqrt{2k_BT\zeta_R}f_n(t),}_\text{Thermal fluctuations} \label{eq:rotorlang}
\end{equation}
where $\zeta_R$ is the effective rotor drag coefficient. Finally, the dynamics of the load are then driven by the motion of the rotor:
\begin{equation}
 \zeta_L\frac{d\theta_L}{dt} = \underbrace{\kappa(\theta_R-\theta_L)}_{\substack{\text{Spring connection}\\\text{to rotor}}} + \underbrace{\sqrt{2k_BT\zeta_L}f_n(t).}_\text{Thermal fluctuations}
\end{equation}
As above, $\zeta_L$ is the effective drag coefficient of the load. 

The above model can be collapsed to explicitly include only the dynamics of a stator with a single loop that generates torque both during its bending ($\phi_S^{i}$ increasing) and unbending ($\phi_S^{i}$ decreasing). This description is isomorphic to the mechanism described previously (see Figure~\ref{fig:mechanism}) because the mechanics of the two half power strokes are equivalent as described above. The equations corresponding to this reduced model are provided in the supplementary text. 

\bibliographystyle{plain}
\bibliography{ref}

\newpage

\begin{center}
 {\LARGE Supplementary Text for}
 
 \setcounter{section}{0}
 \setcounter{equation}{0}

 \vspace{0.5cm}
 {\bf \Large Mechanics of torque generation in the bacterial flagellar motor}
\vspace{0.5cm}

 {\large Kranthi K. Mandadapu*, Jasmine A. Nirody*, Richard M. Berry, George Oster$^{\dagger}$}
 {\normalsize\\ \vspace{0.1cm} *Equal contribution\\ $^\dagger$Corresponding author. Email: goster@berkeley.edu.}
\end{center}

\tableofcontents

\newpage

\section{Model equations}
In the Materials and Methods section in the main text, we provided Langevin equations describing the dynamics of the two stator loops, the rotor, and the load. We explicitly model the dynamics of the stator loops, the rotor, and the load. As described previously, the mechanics of the two stator loops are equivalent. Therefore, a reduced model describing the dynamics of a stator with a single loop that generates torque both during bending ($\phi_S$ increasing) and unbending ($\phi_S$ decreasing) is isomorphic to the one described in the main text. 
\begin{align}
\textbf{Stator : } \hspace{0.5in} \zeta_S\frac{d\phi_S}{dt} &= \underbrace{-\frac{\partial G(\phi_S,j)}{\partial\phi_S}\ell_p}_{\substack{\text{Torque from}\\\text{Proline hinge}}} - \underbrace{\frac{\partial V_{RS}}{\partial\phi_S}}_{\substack{\text{Reaction}\\\text{from rotor}}} -  \underbrace{\frac{\partial\psi}{\partial\phi_S}\ell_p}_{\substack{\text{Electrostatic}\\\text{attraction}}} + \underbrace{\sqrt{2k_BT\zeta_S}f_n(t)}_\text{Thermal fluctuations} \label{eq:statorlang} \\ 
\textbf{Rotor : } \hspace{0.5in} \zeta_R\frac{d\theta_R}{dt} &=  \underbrace{-\frac{\partial V_{RS}}{\partial \theta_R}}_{\substack{\text{Torque}\\\text{from stator}}} - \underbrace{\frac{\partial\psi}{\partial\theta_R}R}_{\substack{\text{Electrostatic}\\\text{attraction}}} - \underbrace{\kappa(\theta_R-\theta_L)}_{\substack{\text{Spring connection}\\\text{to load}}}+ \underbrace{\sqrt{2k_BT\zeta_R}f_n(t)}_\text{Thermal fluctuations} \label{eq:rotorlang}\\
\textbf{Load : } \hspace{0.5in} \zeta_L\frac{d\theta_L}{dt} &= \underbrace{\kappa(\theta_R-\theta_L)}_{\substack{\text{Spring connection}\\\text{to rotor}}} + \underbrace{\sqrt{2k_BT\zeta_L}f_n(t).}_\text{Thermal fluctuations} \label{eq:loadlang}
\end{align}
Here, as before, $\zeta_S$, $\zeta_R$, and $\zeta_L$ are the effective drag coefficients of the stator, rotor, and load. The last term in each equation is the stochastic Brownian force, where $f_n(t)$ is uncorrelated white noise. All other symbols are as described in Table 1 in the main text. In writing the equations of motion, we assume that the stator rotates about the proline hinge. Similarly, the rotor rotates about the axis normal to the plane of the rotor and passing through its center. 

In Equation \eqref{eq:statorlang}, the internal force driving the stator due to the rearrangement of hydrogen bonds caused by a proton binding event is denoted by $F_p = -\frac{\partial G}{\partial\phi_S}$. Here, $G(\phi_S,j)$ denotes the free energy of the stator. Due to the fact that thermal fluctuations are of a comparable magnitude to the free energies considered, the exact shape of the potentials is relatively unimportant. For ease of computation, we approximate the potential using a piecewise linear function. In this setup, the force applied on the rotor by the stator loop is constant and positive during each mechanical power stroke. At other times, there is little elastic strain on the MotA loops, and accordingly, the applied force is near zero.

The torque generated by the stator is dependent on the applied load $\zeta_L$ as a natural consequence of steric forces. A general discussion on contact forces and the explicit formulation of the repulsive interaction potential $V_{RS}$, are provided in Section~\ref{sec:contactforce} and Section~\ref{sec:rotorstatorinteract}, respectively. The contact torque applied to the rotor (in Equation \eqref{eq:rotorlang}), and consequent reaction torque applied to the stator (in Equation \eqref{eq:statorlang}), are given by $\tau_{\text{contact}} = -\frac{\partial V_{RS}}{\partial\theta_R}$ and  $\tau_{\text{reaction}} = -\frac{\partial V_{RS}}{\partial\phi_S}$.

Additionally, the charges on the FliG and the stator loop also exert weak attractive forces on each other. These forces prevent the drifting of the rotor with respect to the stator during the chemical transition events. Further discussion of these forces are included later in Section~\ref{sec:electrostaticsteer}. In the above equations, we denote this term via the term $-\frac{\partial\psi}{\partial\theta_R}R$; it detracts from the repulsive torque imposed by the steric force of the stator.

The rotor and load are connected by a linear spring with constant $\kappa$; the elastic coupling terms in the equations for the rotor and the load thus appear with opposite signs (in Equations \eqref{eq:rotorlang} and \eqref{eq:loadlang}, respectively). The elastic constant in the experiments can vary depending on the length of the hook when attaching the bead. In some cases, the hook is very short or is stiffened by an antibody linker. This corresponds to a large spring coefficient \cite{Block1991, block1989compliance}. An analysis of this model in the corresponding limit $\kappa \to \infty$ is provided in Section~\ref{nospring} of this supplementary text.

\subsection{Simplified deterministic model}
By taking an average over many trajectories, it is possible to generate a deterministic analogue of the model presented above.  Although numerical simulations on the full stochastic model were used for the results in this manuscript, the below formulation is convenient primarily for expository purposes. In particular, it admits explicit analytic solutions for many experimental situations. Before we provide the numerical implementation of the full Langevin equations \eqref{eq:statorlang}-\eqref{eq:loadlang}, we use the following model to introduce several important concepts.

The deterministic equations of motion can be obtained by time averaging the equations \eqref{eq:statorlang}-\eqref{eq:loadlang} as 
\begin{align}
\zeta_S\frac{d\phi_S}{dt} &= F_p\ell_p - \frac{\langle\tau\rangle}{R}\ell_p \label{statlang}\\
\zeta_R\frac{d\theta_R}{dt} &=  \langle\tau\rangle - \kappa(\theta_R-\theta_L) \label{rotlang}\\
\zeta_L\frac{d\theta_L}{dt} &= \kappa(\theta_R-\theta_L). \label{loadlang}
\end{align}
Note that in addition to time averaging, we have neglected the electrostatic term for computational convenience, as it tends to be quite small in value. Here, the average torque on the rotor $\langle\tau\rangle$ results from averaging the torque on the rotor as $\tau_{\text{contact}} = -\frac{\partial V_{RS}}{\partial\theta_R}$ over many trajectories. The return force then can be calculated by $\langle \tau \rangle / R$, which is then multiplied by $\ell_p$ to calculate the return torque. The internal torque of the proline hinge for a given ion motive force (IMF) is $F_p \ell_p = -\dfrac{\partial G}{\partial \phi_S} \propto IMF$. The averaged equations do not contain a noise term because the terms $f_n(t)$ are Gaussian with mean zero. In the following, we compute expressions for the average torque and speed during a single power stroke of the motor from the above deterministic model.

Under the assumption that all active stators act in synchrony, Equation \eqref{statlang} can be generalized to a motor with $N$ stators as follows. An analogue of Equation \eqref{statlang} now corresponds to the motion of the $i$-th stator ($i \in 1,2,\dots N$):
\begin{equation}
\zeta_S\frac{d\phi_S^i}{dt} = F_p\ell_p - \frac{\langle\tau_i\rangle}{R}\ell_p, \ \ \ i \in {1, 2, \ldots, N}.\label{statlangmult} 
\end{equation}
We then can sum the equations of all stators, which results in
\begin{equation}
N\zeta_S\frac{d\phi_S^i}{dt} = NF_p\ell_p - \frac{\langle\tau\rangle}{R}\ell_p.\label{mstatlang}
\end{equation}
Note that we have used the fact that $N \langle\tau_i\rangle = \langle\tau\rangle$. Similarly, an equation for the rotor in a motor with multiple stators can be written as:
\begin{align}
\zeta_R\frac{d\theta_R}{dt} &= \sum_{i=1}^{N} \langle\tau^i\rangle - \kappa(\theta_R-\theta_L)\label{rotlangmult}\\
&= \langle\tau\rangle - \kappa(\theta_R-\theta_L),\nonumber
\end{align}
which is the same form as Equation \eqref{rotlang}. Note that only the terms corresponding to the stators are summed (i.e., the connection term between the rotor and the load is not multiplied by $N$). Equation \eqref{loadlang} also remains as for the single stator case because the spring connection term is unaffected by the addition of torque-generating units.

The three equations of motion \eqref{statlang}-\eqref{loadlang} for a single stator contain 4 unknowns $\{\theta_R,\theta_L,\phi_S,\langle \tau \rangle\}$. This results in an indeterminate system, and requires the addition of an equation to generate a unique solution. This additional equation can be obtained from a fundamental property of contact forces. Since the stator loop is in contact with the rotor during the power stroke, the velocities of the stator loop and the rotor must be equal. This leads to a `contact condition' for the tangential velocities of the stator loop and the rotor:
\begin{equation}\label{contact}
\boxed{\ell_p\frac{d\phi_S}{dt} = R\frac{d\theta_R}{dt}.} \ 
\end{equation}

\subsubsection{Analysis}

Multiplying Equation \eqref{mstatlang} by $\frac{R}{\ell_p}$ and summing with \eqref{rotlang} gives:
\begin{equation}
\frac{NR\zeta_S}{\ell_p}\frac{d\phi_S}{dt} + \zeta_R\frac{d\theta_R}{dt} = NF_pR - \kappa(\theta_R - \theta_L). \label{init}
\end{equation}
Rearranging \eqref{contact}, we get
\begin{equation}
\frac{d\phi_S}{dt} = \frac{R}{\ell_p}\frac{d\theta_R}{dt}.
\end{equation}
Substituting into equation \eqref{init}:
\begin{align}
\frac{NR^2\zeta_S}{\ell_p^2}\frac{d\theta_R}{dt}+ \zeta_R\frac{d\theta_R}{dt} &= NF_pR - \kappa(\theta_R - \theta_L)\nonumber\\
\zeta_R\frac{d\theta_R}{dt} \left(\frac{NR^2\zeta_S}{\ell_p^2\zeta_R} + 1\right) &= NF_pR - \kappa(\theta_R - \theta_L) \nonumber \\
\zeta_R\frac{d\theta_R}{dt} (NM + 1) &= NF_pR - \kappa(\theta_R - \theta_L), \label{eq:filler}
\end{align}
where $M = \frac{R^2\zeta_S}{\ell_p^2\zeta_R}$. Rewriting \eqref{eq:filler} gives us
\begin{equation}\label{eq:filler2}
\frac{d\theta_R}{dt} = \frac{NF_pR}{\zeta_R(NM+1)} - \frac{\kappa}{\zeta_R(NM + 1)}(\theta_R - \theta_L). 
\end{equation}
Dividing equation \eqref{loadlang} by $\zeta_L$ and subtracting \eqref{eq:filler2} leads to
\begin{equation}
\frac{d(\theta_R - \theta_L)}{dt} = \frac{F_pR}{\zeta_R(NM+1)} - \kappa \left(\frac{1}{\zeta_R(NM+1)} + \frac{1}{\zeta_L}\right)(\theta_R - \theta_L).
\end{equation}
This differential equation has the solution:
\begin{equation}
\boxed{(\theta_R - \theta_L)(t) = \frac{A^*(1-e^{-\alpha^*t})}{\alpha^*},} \label{timecourse}
\end{equation}
where $A^* :=  \frac{F_pR}{\zeta_R(NM+1)}$ and $\alpha^* := \kappa \left(\frac{1}{\zeta_R(NM+1)} + \frac{1}{\zeta_L}\right)$.

Substituting expressions for $\dfrac{d\phi_S}{dt}$ and $\dfrac{d\theta_R}{dt}$ from Equations \eqref{mstatlang} and \eqref{rotlang}, respectively, into the equation for contact condition \eqref{contact} leads to:
\begin{equation}
\frac{\ell_p^2}{N\zeta_S}(NF_pR - \tau) = \frac{R^2}{\zeta_R}[\tau - \kappa(\theta_R-\theta_L)].
\end{equation}
Note that, because we are interested in the full time course, we consider $\tau = \tau(t)$, rather than the average torque $\langle\tau\rangle$. Rearranging and solving for $\tau$:
\begin{equation}
\tau = \frac{N\zeta_R R \left[F_p\ell_p^2 - \zeta_SR\kappa(\theta_R - \theta_L)\right]}{\zeta_R\ell_p^2+N\zeta_SR^2}.
\end{equation}
Plugging in the derived expression for $(\theta_R - \theta_L)$ from Equation \eqref{timecourse}:
\begin{equation}
\tau = \frac{F_p\ell_p^2 N R \zeta_R \zeta_S \left[\frac{1}{\zeta_S}+\frac{\left(1-\exp\left[\kappa t \left(\frac{1}{\zeta_L}+\frac{1}{\zeta_R+N R^2 \zeta_S/\ell_p^2}\right)\right]\right) N R^2 \zeta_S}{\zeta_R \left(\ell_p^2 (\zeta_L+\zeta_R)+N R^2 \zeta_S\right)}\right]}{\ell_p^2 \zeta_R+N R^2 \zeta_S}.
\end{equation}

\begin{figure}
\centering
\includegraphics[width=\textwidth]{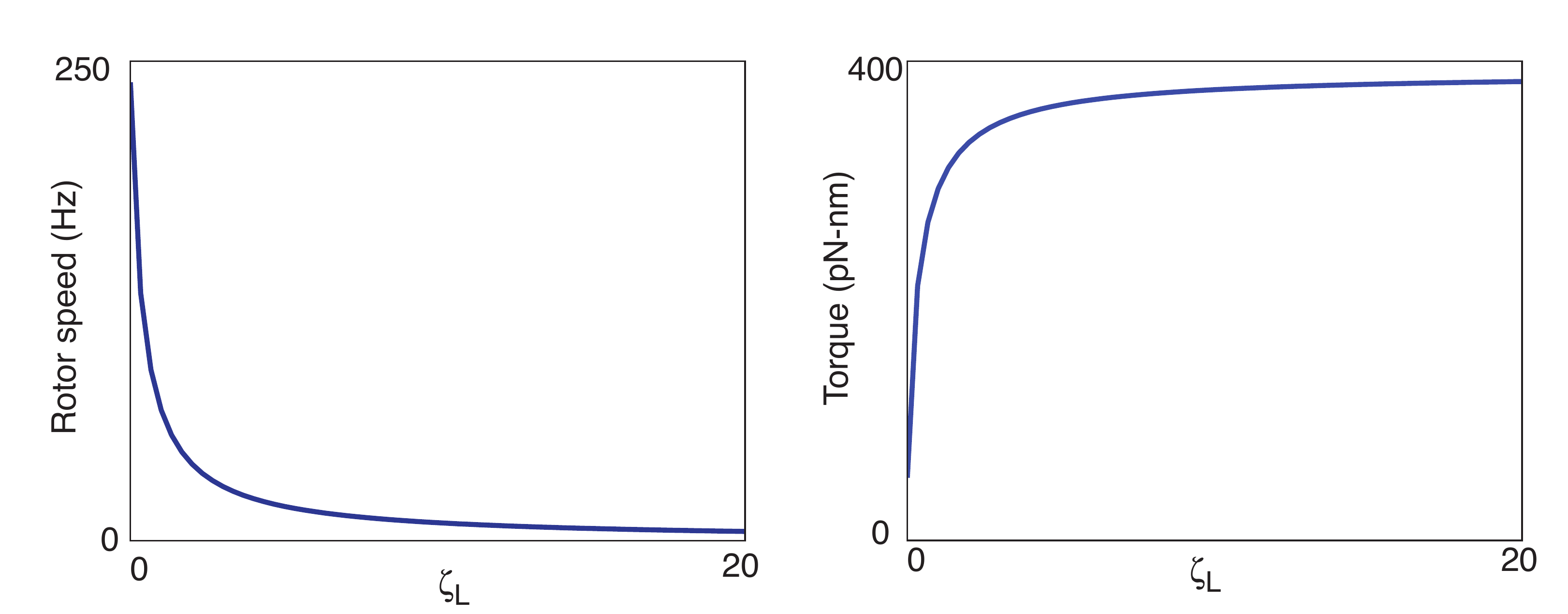}
\label{tsload}
\caption{\textbf{(A)} Rotor speed and \textbf{(B)} Torque as a function of the load drag.  $\zeta_L$. Plotted from Equations \eqref{speed} and \eqref{torque}, respectively.}
\end{figure}

From this value, we can calculate the average torque in a single step $\langle\tau\rangle$ by
\begin{empheq}[box=\widefbox]{align}
\langle\tau\rangle &= \frac{1}{T_m}\int_0^{T_m} \tau(t) dt\nonumber\\
&= \frac{F_p \ell_p^2 N R \left(\left(-1+\exp\left[\kappa T_m \left(-\frac{1}{\zeta_L}-\frac{1}{\zeta_R+N R^2 \zeta_S/\ell_p^2}\right)\right]\right) N R^2 \zeta_L^2 \zeta_S+\kappa T_m (\zeta_L+\zeta_R) \left(\ell_p^2 (\zeta_L+\zeta_R) +N R^2 \zeta_S\right)\right)}{\kappa T_m \left(\ell_p^2 (\zeta_L+\zeta_R)+N R^2 \zeta_S\right)^2}\label{torque}
\end{empheq}
where $T_m$ is the time spent moving during a step. Likewise, we can also calculate the speed of the load $\dfrac{d\theta_L}{dt}$ from Equations \eqref{loadlang} and \eqref{timecourse}:
\begin{align}
\frac{d\theta_L}{dt} &= \frac{1}{\zeta_L}\kappa(\theta_R - \theta_L) \nonumber\\
&= \frac{\left(1-\exp\left[-\kappa t \left(\frac{1}{\zeta_L}+\frac{1}{\zeta_R+N R^2 \zeta_S/\ell_p^2}\right)\right]\right) F_p \ell_p^2 N R}{\ell_p^2 (\zeta_L+\zeta_R)+N R^2 \zeta_S}.
\end{align}
As with torque, we integrate over a time step to find the average speed $\left\langle \dfrac{d\theta_L}{dt}\right\rangle$:

\begin{equation}\label{speed}
\boxed{\left\langle \frac{d\theta_L}{dt}\right\rangle = \frac{F_p \ell_p^2 N R \left[\left(-1+\exp\left[\kappa T_m \left(-\frac{1}{\zeta_L}-\frac{1}{\zeta_R+N R^2 \zeta_S/\ell_p^2}\right)\right]\right) \zeta_L \left(\ell_p^2 \zeta_R+N R^2 \zeta_S\right)+\kappa T_m \left(L^2 (\zeta_L+\zeta_R)+N R^2 \zeta_S\right)\right]}{\kappa T_m \left(\ell_p^2 (\zeta_L+\zeta_R)+N R^2 \zeta_S\right)^2}.}
\end{equation}
Using Equations \eqref{torque} and \eqref{speed}, we can calculate a family of parametric torque-speed curves (parametrized by the load $\zeta_L$), where each curve corresponds to a motor with a constant number of synchronously stepping stators. Computed curves for motors with $1,2,\dots,7$ stators are shown in Figure 7 in the main text.

\subsubsection{Approximation: model without spring}\label{nospring}
In most experimental setups, the filament is removed and a bead is attached to a shortened hook connection. Additionally, the hook is sometimes stiffened with an antibody linker. These setups have a rigid connection between the rotor and the load, corresponding to a large spring constant. In this section, we perform a similar analysis to that in the previous section for the limit $\kappa \rightarrow \infty$. The calculations performed in this section provide analytic formulas for a clear physical understanding of several important properties of the model.

The rotation rates of the rotor and load become equal after an initial ``wind-up'' period. That is the rotor and load move at the same angular speed \Big(i.e., $\dfrac{d\theta_R}{dt} = \dfrac{d\theta_L}{dt}$\Big) after the system reaches a steady state. Note however that the angular positions $\theta_R$ and $\theta_L$ still maintain a (constant) offset. This can be seen explicitly by subtracting Equations \eqref{rotlang} and \eqref{loadlang}:
\begin{align}
\frac{d\theta_R}{dt} - \frac{d\theta_L}{dt} &=  \frac{\langle\tau\rangle}{\zeta_R} - \kappa\left(\frac{1}{\zeta_R}+\frac{1}{\zeta_L}\right)(\theta_R-\theta_L) \nonumber \\
\frac{d(\theta_R - \theta_L)}{dt} &= \frac{\langle\tau\rangle}{\zeta_R} - \kappa\left(\frac{1}{\zeta_R}+\frac{1}{\zeta_L}\right)(\theta_R-\theta_L).\label{steady1}
\end{align}
To simplify some notation, we define $x := (\theta_R - \theta_L)$, $\alpha := \kappa\left(\frac{1}{\zeta_R}+\frac{1}{\zeta_L}\right)$ and $A := \frac{\langle\tau\rangle}{\zeta_R}$, and rewrite Equation~\eqref{steady1}
\begin{equation}\label{dxdt}
\frac{dx}{dt} = A - \alpha x.
\end{equation}
As in the previous section, we can solve for the timecourse $x(t) = (\theta_R - \theta_L)(t)$:
\begin{equation}
\boxed{x(t) = (\theta_R - \theta_L)(t) = \frac{A(1-e^{-\alpha t})}{\alpha}. \label{diffthetas}}
\end{equation}
By definition, $\alpha > 0$, and so it is clear that $x(t)$ will reach a constant value after an initial startup. As $\kappa \to \infty$, the `wind-up' time goes to zero and $\frac{d\theta_R}{dt} = \frac{d\theta_L}{dt}$ in this limit.
\begin{figure}
\centering
\includegraphics[width=\textwidth]{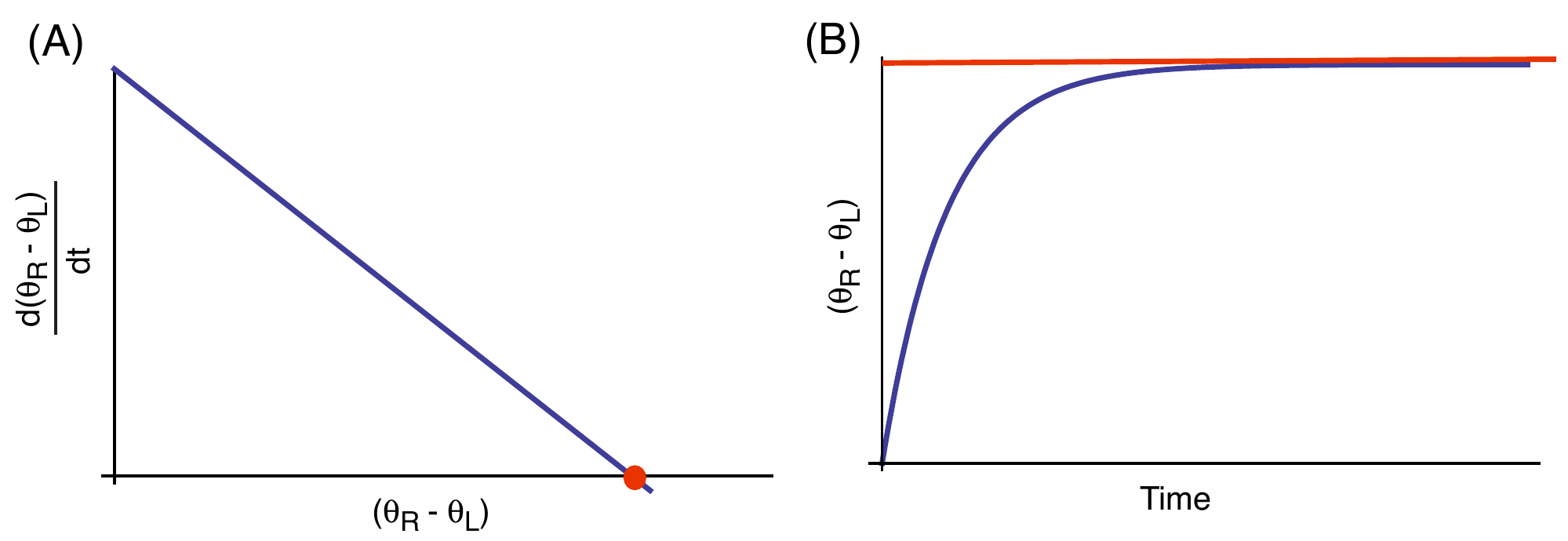}
\label{plotx}
\caption{Angular difference between rotor and load. \textbf{(A)} The change in $(\theta_R - \theta_L)$ with time decreases steadily until a fixed point is reached (marked in red). \textbf{(B)} Accordingly, $(\theta_R - \theta_L)$ increases and asymptotically reaches a steady state. The steady state is denoted by a red line, corresponding to the zero of $\frac{d(\theta_R - \theta_L)}{dt}$, shown as the red dot in \textbf{(A)}.}
\end{figure}

Summing Equations \eqref{rotlang} and \eqref{loadlang} gives us:
\begin{equation}\label{eq:filler3}
\zeta_R\frac{d\theta_R}{dt} + \zeta_L\frac{d\theta_L}{dt} = \langle\tau\rangle.
\end{equation}
We approximate $\frac{d\theta_R}{dt} = \frac{d\theta_L}{dt}$ when $\kappa$ is large, and Equation \eqref{eq:filler3} reduces to:
\begin{equation}
(\zeta_R + \zeta_L)\frac{d\theta_R}{dt} = \langle\tau\rangle. \label{rotloadlang}
\end{equation} 
We multiply Equations \eqref{mstatlang} and \eqref{rotloadlang} by $\ell_p$ and $R$, respectively. After some algebra:
\begin{align}
\ell_p\frac{d\phi_S}{dt} &= \frac{\ell_p^2}{N\zeta_S}\left(NF_p - \frac{\langle\tau\rangle}{R}\right) \label{eq:filler4}\\
R\frac{d\theta_R}{dt} &= \frac{R \langle\tau\rangle}{(\zeta_L + \zeta_R)}. \label{eq:filler5}
\end{align}
Given the relationship \eqref{contact}, we equate the two right hand sides of \eqref{eq:filler4} and \eqref{eq:filler5}:
\begin{align}
\frac{\ell_p^2}{N\zeta_S}\left(NF_p - \frac{\langle\tau\rangle}{R}\right) &= \frac{R \langle\tau\rangle}{(\zeta_L + \zeta_R)}\nonumber\\
\frac{\ell_p^2R}{\zeta_S}\left(NF_pR - \langle\tau\rangle\right) &= \frac{NR^2 \langle\tau\rangle}{(\zeta_L + \zeta_R)}.
\end{align}
Here, the second line is obtained simply by multiplying through by $N$ and $R$. Solving for $\langle\tau\rangle$:
\begin{equation}
\boxed{\langle\tau\rangle = \frac{NF_pR}{\left(1 + \frac{NR^2\zeta_S}{(\zeta_L + \zeta_R)\ell_p^2}\right)}.\label{tau}}
\end{equation}
We can use this expression to attempt some intuition for the result of Sowa \emph{et al.} \cite{sowa2005direct} regarding torque and stator number at high and low loads. Consider the following two limits of Equation \eqref{tau}: (i) high load, when $\zeta_L \gg \zeta_R$, and (ii) low load, where $\zeta_L \ll \zeta_R$. 

In the first case, we have:
\begin{align}
\langle\tau\rangle_{\text{high}} &= \frac{NF_pR}{\left(1 + \frac{NR^2\zeta_S}{(\zeta_L + \zeta_R)\ell_p^2}\right)}\nonumber\\
 & \approx \frac{NF_pR}{\left(1 + \frac{NR^2\zeta_S}{\zeta_L\ell_p^2}\right)}\nonumber\\
 & \approx NF_pR. \label{eq:torque-no}
\end{align}
The third line follows from the fact that $\zeta_S < \zeta_R \ll \zeta_L$. For very high loads, the observed torque is 180 pN-nm, and therefore the force $F_p \approx 9.5$ pN. Equation \eqref{eq:torque-no} suggests that the torque increases linearly with stator number under extremely large loads as observed in the experiments. However, for a given $\zeta_L$, it can be seen that nonlinearities can arise in the torque versus number of stators even in the high load limit. This property is primarily due to the nature of contact forces, and is not applicable to previous models which assume constant torque between stator and rotor. 

Conversely, at low loads ($\zeta_L/\zeta_R \ll 1 $):
\begin{align}
\langle\tau\rangle_{\text{low}} &= \frac{NF_pR}{\left(1 + \frac{NR^2\zeta_S}{(\zeta_L + \zeta_R)\ell_p^2}\right)}\nonumber\\
 & \approx \frac{NF_pR}{\left(1 + \frac{NR^2\zeta_S}{\zeta_R\ell_p^2}\right)}. \label{taulow}
\end{align}

The torque measured at high speeds is approximately 20 pN-nm \cite{lo2013mechanism}. Then, from Equation \eqref{taulow}, the non-dimensional number $NR^2\zeta_S/\zeta_R\ell_p^2 \approx 10$. Also, Equation \eqref{taulow} shows that the torque and speed are not linearly dependent on the number of stators at low loads. This is consistent with previous experimental observations \cite{Yuan2008}. 

The above approximations to the full model are primarily laid out for expository purposes, to introduce general properties of the model. As such, there are several limitations due to the assumptions made, which we outline below. 

Firstly, the observations from Sowa \emph{et al.} are under question given that motors with more than one (or a few) stators may not have been considered in low-load measurements. In order to determine whether the stators do indeed act independently or in synchrony, experiments that directly account for the number of active stators must be performed. The extension of our model to $N$ stators is dependent on the assumption that all stators step in synchrony, and the above results will not hold if stators are independent stochastic steppers. Because of the recent measurements made on single-stator motors, we have chosen to largely focus on the explicit modeling of the intrinsic mechanism of torque-generation in the BFM, rather than on the potential interaction between stators. While a thorough exploration of the dynamics of multiple-stator motors is extremely worthwhile, it is out of the scope of the current paper. 

There are also some inconsistencies in the approximation of an infinitely stiff spring at intermediate and low loads when the motor is not in the mechanically-limited regime. We have that $\ell_p \approx$ 7 nm, $R \approx$ 20 nm, and the non-dimensional number $NR^2\zeta_S/\zeta_R\ell_p^2 \approx 10$ at low loads. Then from Equation \eqref{taulow}, the ratio of the stator and rotor drags $\zeta_S/\zeta_R \approx 1$. Using this, we can estimate the viscosities of the stator and rotor. The drag of the rotor is given by 
\begin{equation}\label{dragrotor}
\zeta_R = \frac{32}{3} \eta_R R^3,
\end{equation}
where $\eta_R$ is the viscosity of the rotor. . An estimate for the drag coefficient of the stator can be obtained by using Equation \eqref{rotloadlang} and the maximum observed speed of the wild-type motor (200 Hz), yielding $\zeta_S = 0.017$ pN-nm-s/rad. Since the drag ratio of the stator and rotor is approximately unity, $\zeta_R \approx 0.017$ pN-nm-s/rad. Using \eqref{dragrotor}, the viscosity of the rotor is approximately $2$ Poise. This is consistent with the fact that the cytoplasm is a mixture of water and proteins \cite{Berg2003}.

Likewise, the drag coefficient of the stator loop with a lever arm of length $\ell_p$ is given by 
\begin{equation}\label{eq:dragstator}
\zeta_S = \frac{\pi}{3} \frac{\eta_S \ell_p^3}{\log(\frac{\ell_p}{R_p}) - 0.66},
\end{equation} 
where $\eta_S$ is the viscosity of the stator and $R_p$ is the radius of the stator loop. Then, using \eqref{eq:dragstator} and a length to width ratio $\ell_p/R_p \approx 10$, we estimate the viscosity of the stator to be 800 Poise. This value is three orders of magnitude higher than the viscosity of a regular lipid membrane and two orders of magnitude higher than the viscosity of a biological membrane.

Moreover, the torque-speed curves produced by the stiff-spring approximation are linear, in contrast to the concave-down torque speed curves observed in single stator motors \cite{lo2013mechanism}. Since the IMF enters only through the dependence of the torque, the speed-IMF curves are also linear at all loads, again in contrast to the recent 100 nm bead experiments on a single stator \cite{lo2013mechanism}.

\subsubsection{Addition of chemical kinetics}
The linear torque-speed curves of a mechanically rate-limiting model elucidate the importance of the inclusion of ion-binding kinetics. In this case, these are events related to the binding of a cation from the periplasm to Asp32 and the unbinding of the cation from Asp32 into the cytoplasm. In this section, we recompute model torque-speed curves explicitly including the dwell times corresponding to the ion-binding and unbinding events between the power strokes.

As done by Meacci and Tu \cite{meacci2009dynamics}, a torque-generation cycle is divided into two parts: (i) moving time $T_m$ and (ii) waiting time $T_w$. Assuming that the ions bind only when the stators are around the minimum of the respective free energy potentials, we may use the above model during $T_m$ and sample $T_w$ from an exponential distribution at the end of each moving step. 

In a motor with a single stator ($N=1$), the instantaneous torque is obtained from Equation \eqref{tau} as 
\begin{equation}
\tau = \frac{F_p R}{\big(1+ \frac{R^2 \zeta_S}{(\zeta_L + \zeta_R)\ell_p^2}\big)}\label{tau2}.
\end{equation} 
The force applied by the proline hinge is given as $F_p = - \dfrac{\Delta G}{\Delta \phi} = -\dfrac{2q\ \text{IMF}}{\Delta \phi}$ , where $q$ is the charge of the ion. The time required to move an angular distance of $\dfrac{2\pi}{26}$ in step $i$ can be calculated from Equation \eqref{tau2} through the relationship $\omega = \tau/(\zeta_R + \zeta_L)$.
\begin{equation}
T_m^i = T_m = \frac{2 \pi }{26} (\zeta_R + \zeta_L )\Big(1+  \frac{R^2 \zeta_S}{(\zeta_L + \zeta_R)\ell_p^2}\Big) \frac{1}{F_p R}.
\end{equation}
Let $T_w^i$ be the waiting time that follows a step $i$. During the waiting time, the instantaneous torque is zero. When the system reaches the steady state, the average torque $\langle \tau \rangle$ can be obtained via a time average
\begin{equation}\label{eq:avgtorque-pmf}
\begin{split}
\langle \tau \rangle & = \lim_{T \to \infty} \frac{1}{T} \int_{0}^{T} \tau(t') \text{d}t' , 
\end{split}
\end{equation}
where $T$ is large. If there are $N$ steps in time $T$, then there are $N$ waiting times. Therefore, Equation \eqref{eq:avgtorque-pmf} can be approximated as 
\begin{equation} \label{eq:avgtorque-pmf-1}
\begin{split}
\langle \tau \rangle  = \lim_{T \to \infty} \frac{1}{T} \int_{0}^{T} \tau(t') \text{d} t' & \approx \displaystyle \frac{ \sum_{i=1}^{N}\tau T_m}{T} \\ 
& = \frac{ \sum_{i=1}^{N}\tau T_m}{\sum_{i=1}^N T_m^i + \sum_{i=1}^N T_w^i} \\
& =  \frac{\tau T_m}{T_m + \frac{1}{N}\sum_{i=1}^N T_w^i} \ . 
\end{split}
\end{equation}
The average dwell time $\frac{1}{N}\sum_{i=1}^N T_w^i$ can be approximated as $\langle T_w \rangle = \left(k_0 \exp\left(\dfrac{IMF}{k_BT}\right)\right)^{-1}$, where $k_0$ is a proportionality constant related to the rate of hopping of the ions. Using this, Equation \eqref{eq:avgtorque-pmf-1} reduces to 
\begin{equation}
\langle \tau \rangle  =  \frac{\tau T_m}{T_m + \langle T_w\rangle}.
\end{equation} 
Likewise, the speed of the rotor (or the bead in the large spring constant limit) can be written as a time average
\begin{equation}\label{eq:avgspeed-pmf-1}
\langle \frac{d\theta_R}{dt} \rangle \approx \lim_{T \to \infty} \frac{1}{T} \int_{0}^{T}   \frac{\text{d} \theta_R(t')}{\text{d} t'} \text{d} t'  = \displaystyle \frac{\frac{2\pi}{26T_m}  T_m}{T_m + \frac{1}{N}\sum_{i=1}^N T_w^i} = \frac{\frac{2\pi}{26T_m}  T_m}{T_m + \langle T_w \rangle}
\end{equation}
where the instantaneous speed is $\frac{2\pi}{26} \times \frac{1}{T_m}$ during the power stroke and zero during the dwell time. 

The torque during a power stroke $\tau$, the time taken by a single step $T_m$ and the dwell time $T_w$ all depend on the IMF. Thus, \eqref{eq:avgtorque-pmf-1} and \eqref{eq:avgspeed-pmf-1} point to the existence of nonlinearity in the torque-speed and speed-pmf curves. These curves are presented and explained in the Results and Predictions section of the main text. 

\section{Overview of steric forces} \label{sec:contactforce}

In this section, we describe our modeling of the steric forces between the stator and the rotor, as well as a general discussion on the nature of contact/steric forces in a low Reynolds number environment. The behavior of objects moving at low Reynolds number is counterintuitive. When the Reynolds number is small, viscous forces dominate over inertial forces and inertia can be ignored \cite{purcell1977life}. In the following, we illustrate some of these properties using a simple linear momentum balance. We can then extend this analysis to angular momentum balances, which are directly relevant to the BFM.

\begin{figure*}[t]
\begin{center}.
\centerline{\includegraphics[width=\textwidth]{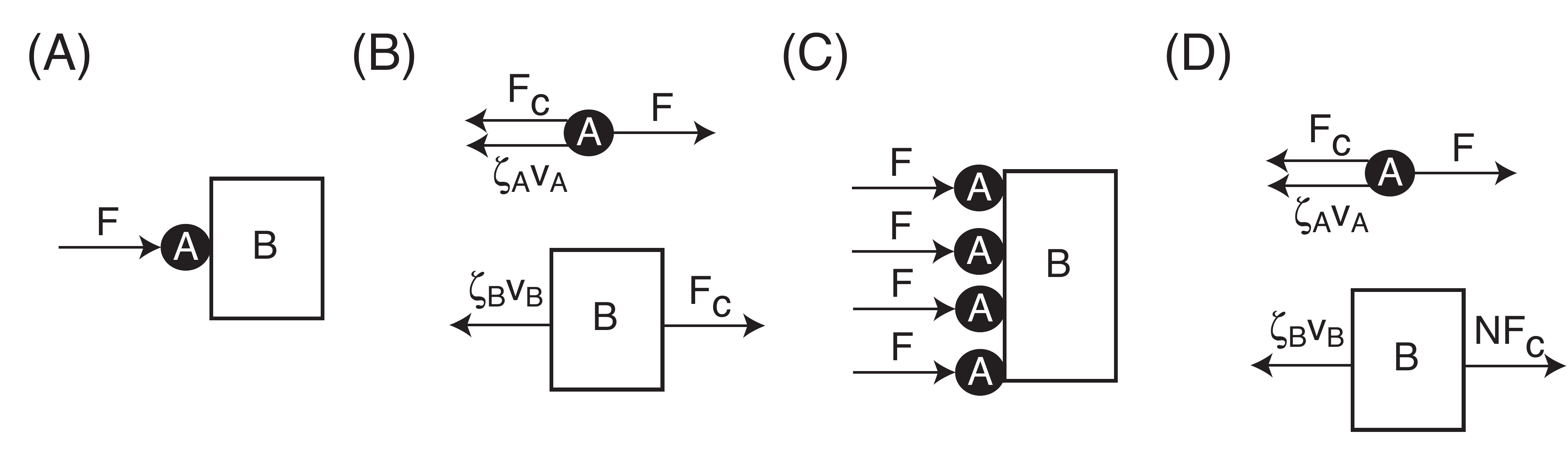}}
\caption{Contact forces. \textbf{(A)} Case 1: a force $F$ is applied to object $A$ which is in contact with object $B$. \textbf{(B)} Free-body diagrams for $A$ and $B$ arranged as in Case 1. \textbf{(C)} Case 2: $N$ objects of type $A$ are in contact with object $B$, each with a force $F$ applied to it. \textbf{(D)} Free-body diagrams for each object of type $A$ as well as $B$ for Case 2.}
\label{fig:contact}
\end{center}
\end{figure*}

Consider a force $F$ pushing an object $A$ which is in contact with a larger object, $B$, as shown in Figure~\ref{fig:contact}. We denote the drag coefficients on objects $A$ and $B$ as $\zeta_A$ and $\zeta_B$, respectively. Let us consider the following two situations.

\textbf{Case 1:} if $F$ is applied to $A$, how much force is transferred to $B$ when they are in contact, as shown in Figure~\ref{fig:contact}A? The corresponding free body diagrams for $A$ and $B$ are shown in Figure~\ref{fig:contact}B; in the steady state, the force balances can be written as:

\begin{equation}\label{eq:linmom_contact}
\begin{split}
F - F_c - \zeta_A v_A & = 0 \  \\ 
F_c - \zeta_B v_B & = 0 \ .
\end{split}
\end{equation}
In Equation \eqref{eq:linmom_contact}, $v_A$ and $v_B$ are the velocities of objects $A$ and $B$, respectively, and $F_c$ is the contact force between objects $A$ and $B$. When the two objects are in contact and are moving together, the following `contact condition' ensures that the velocity of the objects are equal:
\begin{equation}
v_A = v_B.
\end{equation}
In this case, the force transferred by $A$ to $B$ (the contact force, $F_c$), is obtained by solving \eqref{eq:linmom_contact}:
\begin{equation}
F_c = \frac{F}{1+\zeta_A/\zeta_B}.
\end{equation}
When the drag on $B$ is large, \emph{i.e.,} $\zeta_A/\zeta_B \ll 1$, almost all of the force $F$ is transferred to the object $B$. Conversely, when $\zeta_A/\zeta_B  \gg 1$, then most of the force is consumed to drag object $A$ with little force transferred to object B. 

\textbf{Case 2:} when there are $N$ objects of type $A$ in contact with $B$, each with a force $F$ applied to them. This case is shown in Figure \ref{fig:contact}C. As in the first case, we are concerned with how much of $F$ is transferred to $B$.  Again, using free-body diagrams (shown in Figure \ref{fig:contact}D), we can write the equations of motion for the objects as 
\begin{equation}\label{eq:linmom_contact_N}
\begin{split}
F - F_c^i - \zeta_A v_A^i & = 0  \hspace{0.25 in} i \in \{1, \ldots, N\} \ , \\
\sum_{i=1}^{N}F_c^i - \zeta_B v_B & = 0 \ ,
\end{split}
\end{equation}
where $F_c^i$ is the contact force between $i^{th}$ object of type $A$ and object $B$. As before, when the objects are in contact, the contact condition ensures that $v_A^i = v_B$ for all $i$. Therefore, the force transferred to $B$ by $N$ objects of type $A$ can be derived from \eqref{eq:linmom_contact_N} as 
\begin{equation}
\displaystyle F_c =  \sum_{i=1}^{B} F_c^i = \frac{NF}{1+N\zeta_A/\zeta_B}.
\end{equation}
When $\zeta_A/\zeta_B \ll 1$, $F_c \approx NF$. Therefore, the force transferred is multiplied by the number of objects pushing $B$. However, when $\zeta_A/\zeta_B \gg 1$, then the force transferred is $F_c \approx F/(\zeta_A/\zeta_B) \approx 0$. Finally, when $\zeta_A/\zeta_B \approx 1$, then the force transferred is $F_c \approx F$ (i.e. the force transferred in this system is approximately the same as a single object $A$ pushing $B$).

The above properties of contact forces can be applied to the BFM by identifying object $A$ as the stator and $B$ as the rotor-load system (i.e., the rotor and the bead), and $F$ as the internal force generated by the proline hinge pushing the stator from its straight to bent state. 

For large loads (e.g., large beads), almost all of the force generated by the stator is transferred to the rotor; i.e., the torque is close to stall. Moreover, as more stators are recruited, the force transferred increases linearly with the number of active stators, in accordance with the observed linear speed dependence on the number of stators near stall torque. By contrast, at zero load, if there exists a situation where the stator and rotor drags are comparable, then the above analysis suggests that the force transferred during a single step is equivalent to a single stator pushing the rotor. This suggests that, if the assumption that stators step in synchrony holds, torque and speed at low loads may be independent of the number of active stators.

In addition to the above, contact forces also have the following important implications for the BFM: 
\begin{enumerate}
\item Because it operates at low Reynolds number, the rotor moves only as long as it is pushed by the stator. This assures that the rotor never moves faster than the stator. 
\item Experiments based on torque-speed curves alone may never be able to detect the number of operating stators at different loads (e.g., if both torque and speed are independent of the number of stators at low load). Rather, one likely needs chemical markers such as GFP tags as used in \cite{tipping2013load} to identify the number of docked and engaged stators for a given load.
\item The torque generated by the BFM depends on the bead size and is not constant across applied loads, as it was considered to be by previous models  \cite{xing2006torque,meacci2009dynamics,meacci2011dynamics}.  
\end{enumerate}

\section{Numerical implementation}
In the above sections, we have provided the equations of our model, as well as analytic solutions for reduced deterministic approximations. In this section, we provide details on numerical implementation of the full stochastic model, utilizing several arguments from our discussion on steric forces.

\subsection{Interaction potentials}
In this section, we provide explicit forms of the free-energy potentials in terms of the order parameters we choose to describe the motion of the stator and the rotor. 

\subsubsection{Stator potentials}
\begin{figure*}[t]
\begin{center}.
\centerline{\includegraphics[scale=0.7]{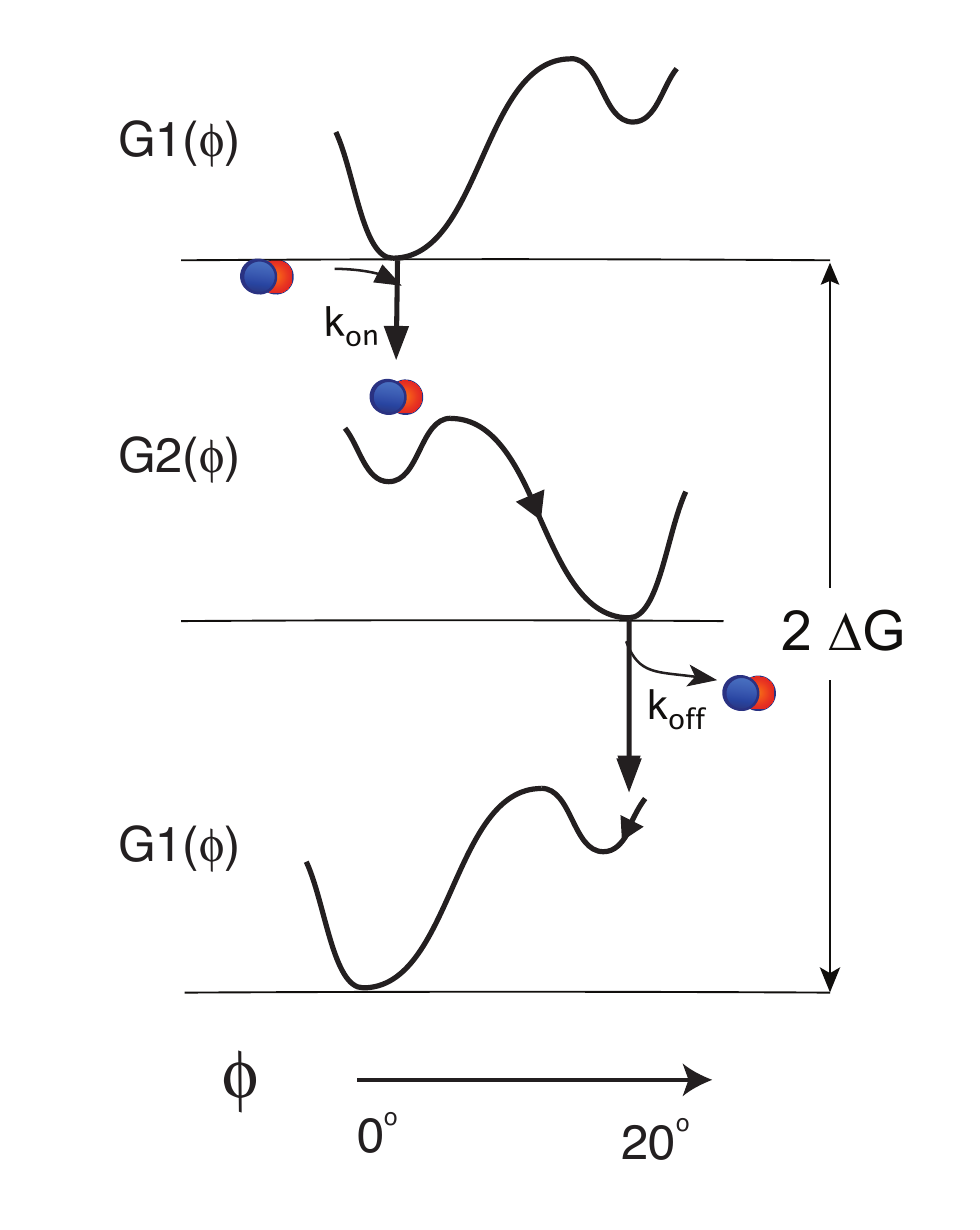}}
\caption{A schematic of the stator potentials.}
\label{fig:g1g2}
\end{center}
\end{figure*}
The order parameter describing the motion of the stator is the angle subtended by the stator loops $\phi^{1,3}_S$ with respect to the vertical MotB ion channels. When two ions bind to the Asp32 residues, the two loops undergo a conformational change from their straight ($\phi_S^{1,3} = 0^{\circ}$) to bent state ($\phi_S^{1,3} = 20^{\circ}$). In this work, we have assumed that the stator loops move in-phase. Therefore, as previously, we model the two stator loop configurations using a single collective parameter $\phi_S$. 

Before the ions bind to the Asp32 residues, the motion of the stators is governed by the potential $G_1(\phi_S)$ where the minimum is around $\phi_S = 0^{\circ}$ as shown in Figure~\ref{fig:g1g2}. When two ions bind to two Asp32s, the stator potential switches from $G_1$ to $G_2$. This compels the stator angle to move from $\phi_S = 0^{\circ}$ to $\phi_S = 20^{\circ}$. 

During this transformation, the loop pushes the rotor via a steric force. At the end of the conformational change, when the loops are at the minimum of the potential $G_2$, the two ions bound to two Asp32s exit into the cytoplasm. The potential then switches back from to $G_1$ and the loops traverse back to $\phi_S = 0^{\circ}$. During this time, the loops apply a contact force on the same FliG as in the previous sub step. 

As noted previously, thermal fluctuations are of the same order of magnitude as the free-energies considered and the precise form of the potentials is not important. In our simulations, we choose:
\[
 G_1(\phi) =
  \begin{cases}
    \beta \phi^2 & \text{if } \phi \leq 0 \\
   F_p \phi       & \text{if } 0 \leq \phi \leq \phi_{max} \\ 
   F_p \phi_{max} + \beta (\phi- \phi_{max})^2 & \text{if } \phi \geq \phi_{max}
  \end{cases}
\]
and 
\[
 G_2(\phi) =
  \begin{cases}
   \beta \phi^2 & \text{if } \phi \leq 0 \\
   -F_p (\phi)       & \text{if } 0 \leq \phi \leq \phi_{max} \\ 
   -F_p \phi_{max} + \beta (\phi- \phi_{max})^2 & \text{if } \phi \geq \phi_{max} .
  \end{cases}
\]

\subsubsection{Rotor-stator interaction potential}\label{sec:rotorstatorinteract}
\begin{figure*}[t]
\begin{center}.
\centerline{\includegraphics[width=0.7\textwidth]{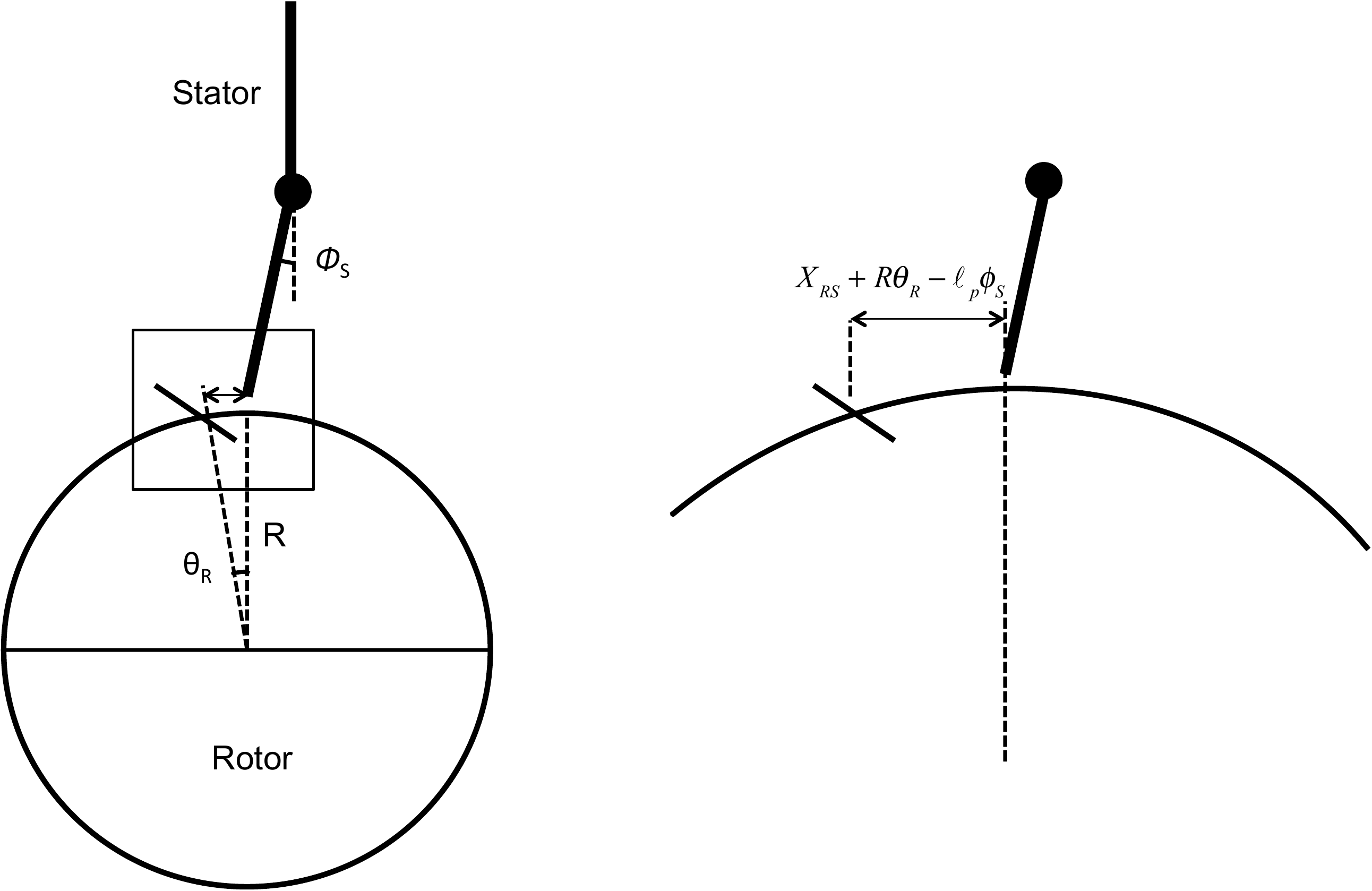}}
\caption{A Schematic of the Geometry of the Rotor-Stator Contact during the half step.
The distance between the nearest FliG and the stator is then given by $x_{RS} = X_{RS} + R\theta_R - \ell_p \phi_S$. The contact force is zero, whenever $x_{RS} \geq X_{RS}$. Moreover, the stator never is ahead of the rotor. This is ensured by choosing a time step such that the fluctuations from the random noise terms are small.}
\label{fig:contactgeo}
\end{center}
\end{figure*}
The steric force between the stator and the rotor can be simulated using a soft linear repulsive force with a cutoff distance $X_{RS}$.  
\[
V_{RS}(\phi_S^{i},\theta_R)  = 
 \begin{cases}
    -F_{RS}\frac{(R\theta_R - \ell_P\phi_S)^2}{X_{RS}} & \text{if } 0 \leq x \leq X_{RS} \\
   0       & \text{otherwise}, 
  \end{cases}
\] 
The torque on the rotor can be obtained as $\tau_{\text{contact}} = -\frac{\partial V_{RS}}{\partial \theta_R}$:
\[
 \tau_{\text{contact}}(x) =
  \begin{cases}
   -F_{RS}R\frac{R\theta_R - \ell_P\phi_S}{X_{RS}} = F_{RS}R\left(1-\frac{x_{RS}}{X_{RS}}\right) & \text{if } 0 \leq x_{RS} \leq X_{RS} \\
   0       & \text{otherwise.}
  \end{cases}
\]
Likewise, the reaction torque on the stator is $-\frac{\partial V_{RS}}{\partial \phi_S}$. $F_{RS}$ is the maximum force that can be applied by the proline hinge and $x_{RS} = X_{RS} + R\theta_R - \ell_p \phi_S$ denotes the distance between the position of the stator loop and the nearest FliG (see Figure~\ref{fig:contactgeo}). All other parameters are defined as described previously.

Note that we assume that weak electrostatic forces place the stator at most 0.5 nm from the nearest FliG on the rotor ($X_{RS} \sim $ 0.5 nm)  prior to the start of the power stroke, as shown in Figure~\ref{fig:contactgeo}. During the half step, the stator moves from $\phi_S = 0^{\circ}$ to $\phi_S = 20^{\circ}$. In this process, a contact force is applied on the rotor when $0\leq x_{RS} \leq X_{RS}$. The contact force is zero when $x_{RS} \geq X_{RS}$. 

All previous theoretical studies of the BFM have chosen to model the interactions between the stator and the rotor as a load-independent force. In the above sections, we have shown that the dependence of steric forces on the load reproduces many of the mechanical characteristic features of the motor found in experiments (e.g., the nonlinearities in the relationship between speed and stator number at high loads). 

While the exact form of the potential used to model the rotor-stator interaction is not very important, we note linear potentials cannot be used. Such potentials result in a constant torque, independent of $\zeta_L$, and thus does not reproduce several properties of steric forces.  

\subsection{Kinetics of ion-binding}

In this section, we will describe how to model ion-binding events as part of the Langevin equation framework. In many problems related to motors, the ion-binding events may occur only in a certain window of a continuous coordinate describing the mechanical motion of the system. In the BFM, there are two main ion-binding events involved in the torque-generation cycle. 
\begin{enumerate} 
\item Two cations (here, protons) from the periplasm bind to the Asp32 residue on two MotBs  or unbind into the cytoplasm when the cytoplasmic loops are straight.
\begin{equation}\label{eq:React_1}
\text {Asp32}^{-} +{\text{H}^{+}_{\text{periplasm}}}\xleftrightarrow[k_{21}]{k_{12}}\text{Asp32-H }
\end{equation}
\item Once the protons are bound, the cytoplasmic loops undergo conformational change from the straight state into the bent state. At the end of the conformational process, the two channels close with respect to the periplasm and instead open towards the cytoplasm. Once this occurs, the two protons unbind from Asp32 residues into the cytoplasm (or, in the reverse reaction, bind from cytoplasm to Asp32). 
\begin{equation}\label{eq:React_2}
\text{Asp32-H } \xleftrightarrow[\tilde{k}_{12} ]{\tilde{k}_{\mathrm{21}}}\text {Asp32}^{-} +{\text{H}^{+}_{\text{cytoplasm}}} 
\end{equation}
\end{enumerate}    
First, we will describe the case for the kinetics at equilibrium (i.e., at zero IMF), followed by the procedure to model the kinetics under a non-zero IMF. 

\subsubsection{Equilibrium kinetics under zero IMF} 
In the following we consider the proton-driven motor of \emph{E. coli}, and so the IMF concerned is the proton motive force (PMF). When there is no PMF, the forward and backward reaction rates for the reaction in \eqref{eq:React_1} must satisfy
\begin{equation}
 \frac{k_{12}}{k_{21}} = \frac{[\text{Asp32-H}]}{[\text{Asp32}] } \ .
\end{equation}
Let the equilibrium dissociation constant of reaction \eqref{eq:React_1} be defined as 
\begin{equation}
\text{K}_\text{a}^p = \frac{\text{[Asp32][H}^{+}_{\text{periplasm}}] } {[\text{Asp32-H}]} \ .  
\end{equation}
If $\text{pK}_\text{a}^p = -\log_{10}\text{K}_\text{a}^p$, then the forward and backward rates should satisfy 
\begin{equation}
 \frac{k_{12}}{k_{21}}  =  10^{({\text{pK}_\text{a}^p} - \text{pH}_{\text{periplasm}})} \ .
\end{equation}
Note that the acid dissociation constant value of $\text{pK}_\text{a}^{p}$ should be determined from experiments. A similar relation to the above can be derived for the reaction in \eqref{eq:React_2}.

\subsubsection{Kinetics under non-zero IMF}
To satisfy detailed balance when the PMF is non-zero, the kinetic coefficients for the reaction should satisfy 
\begin{equation}
 \frac{k_{12}}{k_{21}}  =  \displaystyle 10^{({\text{pK}_\text{a}^p} - \text{pH}_{\text{periplasm}})} \exp\bigg(\frac{e\psi_p + G_1 - G_2}  {k_B T}\bigg) \ .
\end{equation}

There exist multiple choices for the expressions $k_{12}$ and $k_{21}$. For convenience and symmetry, we choose the following for the rate constants : 

\begin{equation}
k_{12} =  \displaystyle 10^{- \text{pH}_{\text{periplasm}}} \exp\bigg(\alpha\Big(\frac{e\psi_p + G_1 - G_2}  {k_B T}\Big)\bigg) \ ,
\end{equation}

\begin{equation}
{k_{21}}  =  \displaystyle 10^{{-\text{pK}_\text{a}^p}} \exp\bigg((1+\alpha)\Big(\frac{e\psi_p + G_1 - G_2}  {k_B T}\Big)\bigg)  \ .
\end{equation}

\section{Electrostatic steering}\label{sec:electrostaticsteer}

In this section, we provide more detailed calculations for our electrostatic steering hypothesis, as shown in the main text. We also provide support for the dipole approximation via comparison with an analogous calculation with point charges. 

We note that, because no structure of the stator is yet available, the purpose of these calculations is quite qualitative. Our goal is to simply to predict what the electrostatic energy landscape should look like in order to support our steering hypothesis.

\subsection{Dipole approximation}

As a first-order approximation, we consider the relevant charges on the FliG helix and the stator loops to be dipoles. We denote the rotor dipole moment as $\vec{p}_R^{\ k}$, where $k \in \{1\dots 26\}$ enumerates the number of FliGs on the rotor periphery. Likewise, the stator dipole moment is denoted as $\vec{p}_S^{\ i,N}$, where $N$ enumerates the number of stator units, and $i$ the number of loops on a single stator. Here for ease of exposition, we show electrostatic calculations for a single stator with a single loop ($i = 1$ and $N=1$), but note that this calculation can be easily extended to the full stator model.

We calculate the electric field felt by the stator loop as:
\begin{equation}
E = \sum_{k = 1}^{26} E_k,
\end{equation}
where
\begin{equation}
E_k = \frac{1}{4\pi\epsilon |r_k|^3}\left[3(\vec{p}_R^{\ k}\cdot\hat{r}_k)\hat{r}_k - \vec{p}_R^{\ k}\right].
\end{equation}
Here $\epsilon$ is relative permittivity of cytoplasm and $\vec{r}_k$ is a vector quantity which denotes the distance between the stator loop and the $k$th FliG.

From $E_k$, we calculate the interaction energy between the dipole on the stator loop $\vec{p}_S$ and the $k$th FliG $\vec{p}_R^{\ k}$ as:
\begin{align}
U_k &= -\vec{p}_S \cdot E_k\\\nonumber
&= \frac{1}{4\pi\epsilon |r_k|^3}\left[(\vec{p}_R^{\ k} \cdot \vec{p}_S) - 3(\vec{p}_R^{\ k} \cdot \hat{r}_k)(\vec{p}_S\cdot\hat{r}_k)\right].
\end{align}
Similar to the calculation for total energy, we have
\begin{equation}
U_{\text{stator}} = \sum_{k=1}^{26} U_k.
\end{equation}
Note that, as the distance from the stator loop increases, the terms in the total energy sum drop off as $\frac{1}{|r|^3}$, and so the contribution by FliGs located far from the stator loop is not appreciable.

To calculate the total energy in the system, we also add in the interaction energies between pairs of FliG molecules. However, we note that their relative positions do not change as the stator rotates, and so this consideration results simply in a translation of the entire landscape and has no effect on the topology.
\begin{equation}
U_{\text{total}} = U_{\text{stator}} + \sum_{i=1}^{26} \sum_{j=1, j\neq i }^{26} \frac{1}{4\pi\epsilon |r_{i,j}|^3}\left[(\vec{p}_R^{\ i} \cdot \vec{p}_R^{\ j}) - 3(\vec{p}_R^{\ i} \cdot \hat{r}_{i,j})(\vec{p}_R^{\ j}\cdot\hat{r}_{i,j})\right],
\end{equation}
where $r_{i,j}$ is the displacement vector between the $i$th and $j$th FliG.

\subsection{Comparison with calculation using point charges}

We also performed the above calculations using point charges for the relevant residues on FliG and MotA loops. Because of the uncertainty in the position of the charges, we are interested primarily in a qualitative affirmation of our approximation --- that is, the existence of a gently sloping, relatively wide energy well. 

As in our dipole calculations, we consider the charged residues on the FliG to be positioned at $\frac{\pi}{4}$ to the horizontal (i.e., the charges are positioned along the dipole as shown in Figure 3 in the Main Text). Similarly, we position the charges on the stator loop along the dipole, positioned radially outward from the stator center. Given this configuration, we calculate the electrostatic energy as
\begin{align}
U_{\text{pc}} &= \underbrace{\sum_{i=1}^{8} \sum_{j=1, j\neq i}^{8} \frac{q_i q_j}{4\pi\epsilon |r_{i,j}|}e^{-|r_{i,j}|/\lambda_D}}_{\text{Interaction between stator charges}} +  \underbrace{\sum_{i=1}^{52} \sum_{j=1, j\neq i }^{52}  \frac{q_i q_j}{4\pi\epsilon |r_{i,j}|}e^{-|r_{i,j}|/\lambda_D}}_{\text{Interaction between FliG charges}}\\\nonumber  
& + \underbrace{\sum_{i=1}^{8} \sum_{j=1}^{52} \frac{q_i q_j}{4\pi\epsilon |r_{i,j}|}e^{-|r_{i,j}|/\lambda_D}.}_{\text{Interaction between charges on stator and FliGs}}
\end{align}

We screen charges using a Debye length of $\lambda_D = 0.5$ nm. As before, $r_{i,j}$ denotes the displacement vector between charge $i$ and charge $j$. Note that, similar to our dipole calculations, the first two terms of the energy are invariant to rotation of the stator. Therefore, they only serve to translate the entire energy landscape and do not affect the topology. 

\begin{figure}
\centering
\includegraphics[width=\textwidth]{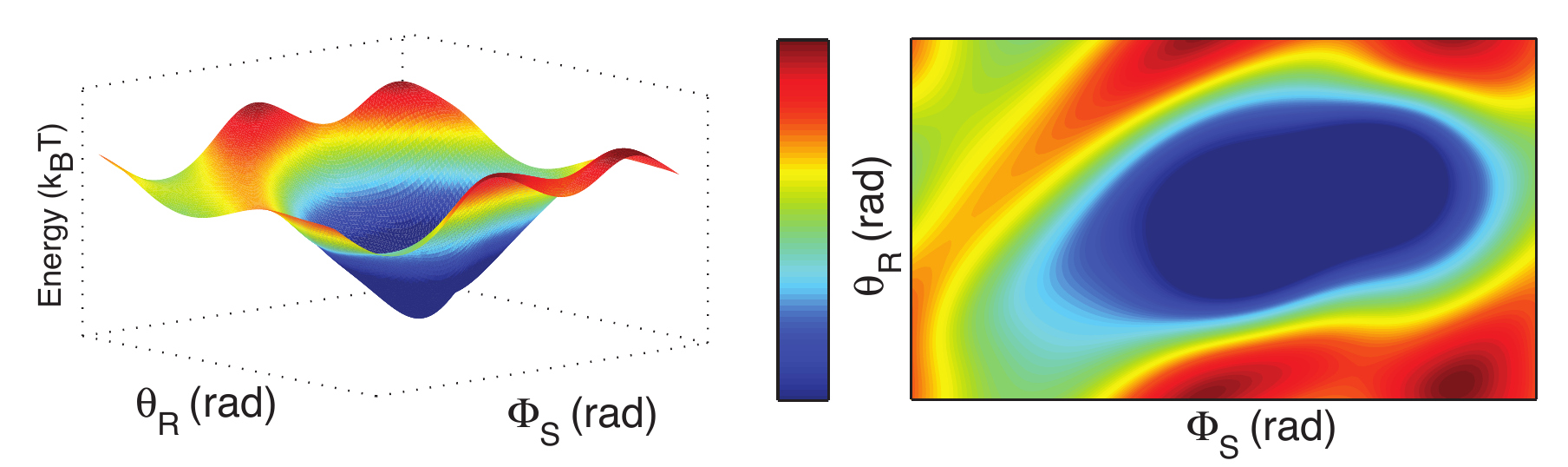}
\caption{Energy landscape during `electrostatic steering' calculated using point charges rather than dipoles. All parameters are defined as in the main text; the stator is centered at (21,-2,1) and the rotor is centered at the origin (all distances in nm).}
\label{electro}
\end{figure}

As shown in Figure \ref{electro}, the topology generated is indeed similar to that using the dipole approximation in the main text. The removal of certain charges may modify the energy landscape via a flattening or widening of the energy well. In conjugation our electrostatic steering hypothesis, this may point to a possible explanation for experimental studies which show that mutation of charged residues reduce, but do not eliminate, motor function. Our methods can easily be extended using positional information from solved structures to more quantitatively analyze the effects of these charges. While such an investigation is certainly warranted, it is out of the scope of the current paper.

\subsection{Centering the stator}

There is little structural information for the stator of the BFM. Therefore, in our calculations, we choose a possible position of the stator which is consistent with our model and experimental results. Firstly, we center the stator so that all four stator loops are able to access the FliG molecules. The occasionally observed backsteps imply that both pairs of loops must be able to execute a power stroke during motor function.
 
Preliminary calculations strongly suggest that the loop charges cannot be in plane with the charges on the FliG. The close proximity of the charges results in an electrostatic interaction far too strong to be feasible given the known efficiency of the BFM (i.e., far too much energy would be wasted in ``letting go''). In-plane configurations which do not present both extremely steep and deep wells must place the stator out of reach of the FliGs. 

Given these considerations, and noting that the stator and rotor have radii of respective lengths $\sim$20~nm and $\sim$2 nm, we center the stator at $(x_S,y_S,z_S)=(21,-2,1)$ using a coordinate system with the rotor centered at the origin. 

\section{An Ising model to describe conformational changes in FliG}

The existence of motor backsteps in the absence of CheY-P has been largely attributed to ``microscopic reversibility''. Given this constraint, there are three possibilities in explaining the backstep: (i) A backward transit of an ion from the cytoplasm to the periplasm. (ii) A sufficiently strained hook relaxes by moving the rotor backwards. (iii) A fluctuation which changes the orientation of FliG from its CCW to CW orientation. 

The first two possibilities contradict our proposed model. For example, if the first case were to be correct, then a backstep would correspond to a reversal of the entire conformational change process. In this case, an ion bound to the Asp32 from the cytoplasm should restore the MotA helix from its bent state to the straight state in the presence of the ion. This is in direct contradiction with one of our model assumptions that MotA relaxes to the bent state due to the rearrangement in the hydrogen bonds caused by the binding of the ion to the Asp32 residue. 

The second possibility (that the relaxation of an elastically strained hook due to several sequential forward steps leads to an occasional backstep) is also not feasible within the context of our model: a contact force guarantees that the rotor always follows the stator. 

In this work, we attribute the molecular basis for the backstep to the third possible scenario: a conformational change in one (or more) individual FliGs on the periphery of the rotor. A FliG can exist in two states; in our model, these states correspond to two orientations of the FliG dipole vector.  Our electro-steric model, in conjunction with an Ising model corresponding to the states of FliG, explains the existence of occasional backsteps in a self-consistent manner. 

Briefly, a backstep results from a fluctuation of FliG from the CCW to the CW position.  In this case, when the motor is predominantly moving in the CCW direction and whenever a FliG changes its state from CCW state to CW state and is close to a stator, then the stator---using MotA loops 2 and 4---applies the contact force and pushes the FliG in the backward direction. 

While the configuration of the FliGs in the CW orientation is fairly well agreed upon, the exact CCW configuration is still under debate. At least three possible  orientations have been suggested, including directions orthogonal or $180^{\circ}$ with respect to the CW orientation. Despite this uncertainty, it is generally believed that there are two significantly different orientations for each FliG. 

As in other models, we describe the transitions between the two FliG orientations by a one-dimensional periodic Ising model consisting of 26 spins (corresponding to 26 FliGs). Each individual spin $s_i$ can exist in two possible states corresponding to the two orientations of FliGs: $s_i = +1$ (CCW) and $s_i = -1$ (CW). The Hamiltonian is 
\begin{equation}\label{eq:ising}
H = -J\sum_{i,j=1}^{N} s_i s_j - h \sum_{i=1}^{N} s_i
\end{equation}  
where $J$ denotes the nearest neighbor pairwise interaction energy and $h$ denotes the field biasing the FliGs to preferentially orient in a certain direction. Let $\{s_i\} = (s_1, s_2, s_3, \dots, s_{26})$ denote a possible state of the rotor. The probability of such a state is given by the Boltzmann distribution $\dfrac{e^{-\beta H(\{s_i\})}}{Z(J,H,\beta)}$, where $Z(J,H,\beta)$ is the partition function. The partition function for a one-dimensional Ising model \eqref{eq:ising} is obtained exactly using the transfer-matrix approach, and is given by
\begin{equation}
Z(J,h,\beta) = \lambda_+^{N} + \lambda_-^N   
\end{equation}
where 
\begin{equation}
\lambda_{\pm} = e^{\beta J} \cosh(\beta h) \pm \sqrt{e^{2\beta J} \sinh^2(\beta h) + e^{-2\beta J}} \ .
\end{equation}

In order to understand the back steps while the motor is running predominantly in CCW direction, we calculate the probability that any FliG (or any spin) points in the CW direction. This is given by 
\begin{equation}
p({s_i = -1})  =  \sum_{(s_1,\ldots,s_{i-1},-1,s_{i+1},\ldots,s_{26}) } \frac{e^{-\beta H((s_1,\ldots,s_{i-1},-1,s_{i+1},\ldots,s_{26}))}}{Z}
\end{equation}
where the summation is over all possible states fixing $s_i = -1$. In the following, we derive an analytical expression for the probability $p({s_i = -1})$. 

Let us denote the probability for the spin $s_i = +1$ as $p({s_i = +1})$. Given the probabilities $p({s_i = +1})$ and $p({s_i = -1})$, the average value $\langle s_i \rangle$ of the spin $s_i$ can be calculated as 
\begin{equation}\label{eq:mean_spin}
\langle s_i \rangle = (1)p({s_i = +1})  + (-1)p({s_i = -1})   = 1-2p({s_i = -1}).
\end{equation}
Moreover, the derivative of the partition function $Z$ with respect to the field $h$ yields 
\begin{equation}\label{eq:mean_magnet}
\frac{1}{N\beta Z }\frac{\partial Z}{\partial h} = \frac{1}{N}\sum_{\{s_i\}} \sum_i s_i  \frac{e^{-\beta H(\{s_i\})}}{Z} = \langle s_i \rangle \ . 
\end{equation}
Therefore, using \eqref{eq:mean_spin} and \eqref{eq:mean_magnet}, the probability that spin $s_i = -1$ can be obtained as 
\begin{equation}\label{eq:prob_backstep}
p({s_i = -1}) = \frac{1 - \frac{1}{N\beta Z }\frac{\partial Z}{\partial h} }{2}  \ .
\end{equation}
The derivative of the partition function with respect to the field $h$ can be evaluated using the derivatives of the $\lambda_{\pm}$ functions, which are given by 
\begin{equation}
\frac{\partial \lambda_{\pm}}{\partial h} = \beta \sinh(\beta h) \Bigg(e^{\beta J} \pm \frac{e^{2\beta J} \cosh(\beta h)}{\sqrt{e^{2\beta J} \sinh^{2}(\beta h) + e^{-2\beta J}}} \Bigg) \ .
\end{equation}

Within the context of our model, the likelihood of a flip in FliG conformation corresponds directly to the likelihood of a backstep. That is, the probability that a FliG is in the CW state is given by \eqref{eq:prob_backstep}. This is because the fundamental mechanics of a backstep is the same as that of a forward step; the difference between these two scenarios is wholly described by the difference in FliG configuration. 

Suppose that the motor is moving primarily in the CCW direction. This means that the FliGs in close proximity to the stator loops are oriented to favor CCW rotation (that is, to favor interaction with MotA loops 1 and 3). However, if a FliG close to the stator is oriented to favor CW rotation, then the FliG interacts with MotA loops 2 and 4, resulting in a step in the CW direction (a ``backstep'' when the motor is moving primarily CCW). This probability is given by \eqref{eq:prob_backstep}.

\begin{figure}[h!]
\includegraphics[width=\textwidth]{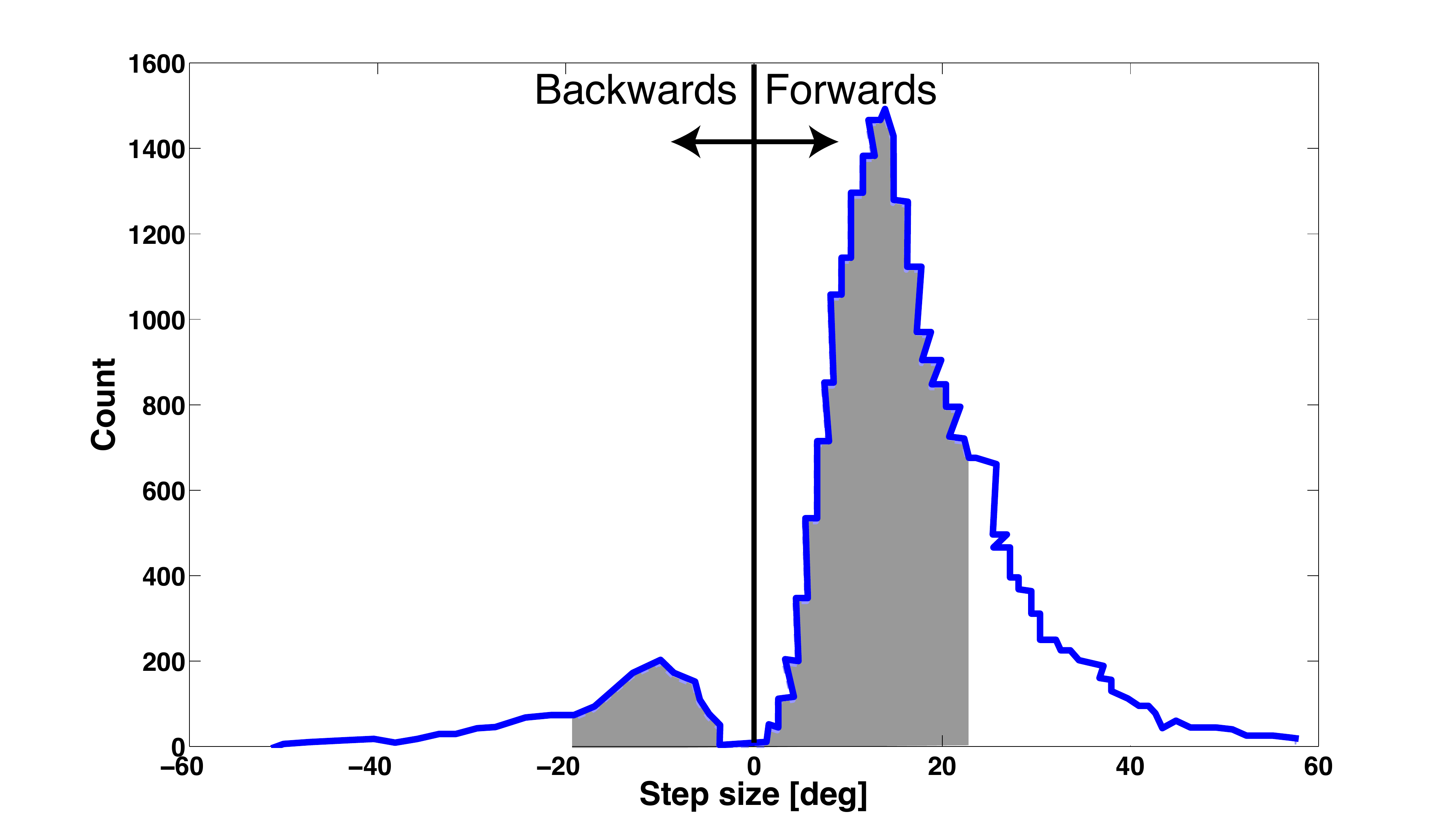}
\caption{Probabilities of backward steps relative to forward steps can be estimated using statistics from stepping experiments (data shown from \cite{sowa2005direct}). The area under the curve for steps below and above 0$^{\circ}$ corresponds to forward and backward step probabilities, respectively. The areas on either side are calculated as highlighted in gray. Curves are truncated where each shows a second ``peak'', corresponding to two steps being blurred together due to limitations in experimental resolution. The relative areas calculated for the data above suggest that the probability of a backwards step is 7.3\%.}
\label{fig:steps}
\end{figure}

For the choice of the energy scale $\beta J = 2$ at room temperature and a biasing field of $\beta h = 0.05$, the probability that any FliG is in the CW state is $p({s_i = -1}) = 0.08$. That is, on average, 8 out of every 100 torque-generating cycles will result in a backstep. Areas under the curve corresponding to forward and backward steps from data collected by Sowa \emph{et al.} \cite{sowa2005direct} indicate that $p({s_i = -1}) \sim$ 0.073 (Figure \ref{fig:steps}).
 
The above analysis provides an explanation for how backsteps in the absence of CheY-P can arise from fluctuations in FliG configurations. However, it does not take into account how the above probabilities are effected by load or PMF. For example, the timescale of a single step depends on the load. If one assumes that this step is a backstep, then this particular FliG is pinned in the ``backwards'' orientation for the duration of that step. This CW defect in the FliG ring can affect the switching probabilities of the neighboring FliGs, resulting in further defects along the ring. Therefore, at higher loads, the probability of two (or more) subsequent backsteps may not be negligible. However, a complete understanding of the above requires a far more detailed analysis of an Ising model in conjunction with the proposed electrosteric model than is within the scope of this work. 

\newpage

\bibliography{ref}

\end{document}